\documentclass[
preprint, nofootinbib, aps, prd, superscriptaddress
]{revtex4-1}

\usepackage{bm}
\usepackage{physics, amsmath, amssymb, amsfonts, siunitx, tensor, paralist, comment, setspace}
\usepackage[font=small,labelfont=bf,justification=justified, singlelinecheck = false]{subcaption}
\usepackage[font=small,labelfont=bf,justification=Justified, singlelinecheck = false]{caption}
\usepackage{graphicx, color}
\usepackage{float} 
\usepackage{setspace}
\usepackage{hyperref}
\usepackage[mathlines]{lineno}

\hypersetup{%
bookmarksnumbered=true,%
bookmarksopen=true,%
colorlinks=true,%
linkcolor=blue,
citecolor=red,
}

\allowdisplaybreaks[4]

\newcommand\nn{\nonumber \\}
\newcommand\e{\mathrm{e}}

\newcommand{\figref}[1]{Fig.~\ref{#1}}

\newcommand{\secref}[1]{Sec.~\ref{#1}}

\begin{document}
\count\footins = 1000

\preprint{KEK-TH-2646, KEK-Cosmo-0354}

\title{Compact star in noninteger power model of \texorpdfstring{$f(R)$}{TEXT} gravity}%

\author{Yong-Xiang~Cui}
\email{cyx2021112305@mails.ccnu.edu.cn}
\affiliation{
Institute of Astrophysics, Central China Normal University, Wuhan 430079, China
}

\author{Zu~Yan}
\email{yanzu0015@gmail.com}
\affiliation{
Institute of Astrophysics, Central China Normal University, Wuhan 430079, China
}

\author{Kota~Numajiri}
\email{numajiri.kota.m3@s.mail.nagoya-u.ac.jp}
\affiliation{
Department of Physics, Nagoya University, Nagoya 464-8602, Japan
}

\author{Taishi~Katsuragawa}
\email{taishi@ccnu.edu.cn}
\affiliation{
Institute of Astrophysics, Central China Normal University, Wuhan 430079, China
}

\author{Shin'ichi~Nojiri}
\email{nojiri@gravity.phys.nagoya-u.ac.jp}
\affiliation{
KEK Theory Center, 
High Energy Accelerator Research Organization (KEK), Oho 1-1, Tsukuba, Ibaraki 305-0801, Japan
}
\affiliation{
Kobayashi-Maskawa Institute for the Origin of Particles and the Universe, 
Nagoya University, Nagoya 464-8602, Japan
}

\begin{abstract}
We investigate compact stars in the noninteger power (NIP) model of $f(R)$ gravity theory, which includes the higher-curvature correction to the Einstein-Hilbert action.
The mass-radius relation of the compact stars in the NIP model predicts large deviations from those in the general relativity in the low-mass region, potentially allowing us to test the NIP model by future astrophysical observations.
We also study the nonvanishing scalar hair surrounding the compact star and demonstrate that the chameleon mechanism works efficiently. 
They result in distinct scalar profiles inside and outside the star, which implies screening the fifth force mediated by the scalar field.
\end{abstract}

\maketitle
\tableofcontents


\section{Introduction}

Many independent observations have shown that the current Universe is expanding at an accelerating rate~\cite{SupernovaSearchTeam:1998fmf}. 
In order to explain the accelerated expansion, we inevitably introduce dark energy (DE)~\cite{Tsujikawa:2008zz, Caldwell:2009zzb, Bamba:2012cp, Lobo:2008sg, Huterer:2017buf}, a new source of energy that provides negative pressure and repulsive force. 
The $\Lambda$ cold-dark-matter (CDM) model~\cite{Planck:2018vyg, Bull:2015stt} provides one of the straightforward explanations for DE, where DE is interpreted as the cosmological constant in general relativity (GR).
This standard model of cosmology succeeds in explaining almost the entire cosmic history. 
However, it also suffers from the so-called cosmological constant problem, i.e., why the observed cosmological constant is so small and why it is of the order of the critical density of the current Universe.

Modified gravity theory introduces a new action instead of the Einstein-Hilbert action, which will lead to the emergence of a new degree of freedom expressed as a new dynamical field~\cite{Copeland:2006wr, Nojiri:2010wj, Clifton:2011jh, Nojiri:2017ncd}. 
If the new dynamical field can act as DE, then we could explain the accelerated expansion of the late-time Universe without introducing the cosmological constant~\cite{Nojiri:2006ri, Bernardo:2022cck, Nojiri:2003ft}.
The above motivates us to investigate the late-time cosmological models of modified gravitational theories.
Moreover, the Hubble tension problem has gained increasing attention in recent years~\cite{ Sakstein:2019fmf, Niedermann:2020dwg, Vagnozzi:2021gjh}. 
The value of the Hubble constant estimated by the local measurements in the late-time Universe is larger than that inferred from observational data in the early Universe based on the $\Lambda$CDM model.
The cosmological models predicted in the modified gravity theory may allow us to tackle the Hubble tension~\cite{DiValentino:2021izs, Odintsov:2020qzd, Chen:2024wqc}.
The latest Dark Energy Spectroscopic Instrument result~\cite{DESI:2024mwx} also supports the cosmological models beyond the $\Lambda$CDM.

$f(R)$ gravity theory is one of the modified gravity theories~\cite{Sotiriou:2008rp,DeFelice:2010aj}, 
where the Einstein-Hilbert action is replaced by the function of the scalar curvature $R$, $f(R)$.
Such a modification introduces a new degree of freedom expressed as a scalar field~\cite{Capozziello:2011wg}, the scalaron.
The scalaron field can drive the accelerated expansion of the Universe, and $f(R)$ gravity theory has been investigated in the context of the dynamical DE~\cite{Starobinsky:2007hu, Hu:2007nk, Tsujikawa:2007xu, Elizalde:2010ts} and the inflaton ~\cite{Starobinsky:1980te, Pi:2017gih, Wang:2024vfv}.
Although the Lovelock theorem states that GR is a unique metric theory whose equation of motion includes up to the second-order derivatives in four-dimensional spacetime~\cite{Lovelock:1971yv, Lovelock:1972vz}, 
$f(R)$ gravity includes the derivatives of the metric higher than the second order in the field equations, which violates assumptions in the Lovelock theorem.

In study of DE models of $f(R)$ gravity theory, their functional forms of $f(R)$ are constructed from phenomenological perspectives,
while they are also constrained by demanding theoretical consistency, such as the absence of antigravity and stability under the perturbation.
The scalaron field shows up as an additional degree of freedom (DOF) from the gravity sector.
Thus, if the scalaron field plays a role of dynamical DE at low-energy scales,
DE models of $f(R)$ gravity include a parameter corresponding to the cosmological constant.
On the other hand, DE models of $f(R)$ gravity suffer from the curvature singularity problem~\cite{Frolov:2008uf, Dutta:2015nga, Miranda:2009rs} at high-energy scales.
It is known that adding the higher-curvature correction can cure the curvature singularity~\cite{Appleby:2009uf}, and the higher-curvature term can dominate at scales larger than DE, such as the astrophysical scale.
From a phenomenological viewpoint, it is significant to examine the higher-curvature corrected models of $f(R)$ gravity theory.

Studying compact stars allows us to test the gravitational theory in a strong and nonperturbative gravitational-field environment~\cite{Perrodin:2017bxr}.
Compact stars are in the hydrostatic equilibrium between internal matter and gravity, which is described by the Tolman-Oppenheimer-Volkoff (TOV) equation in GR~\cite{Chandrasekhar:1931ih, Tolman:1939jz, Oppenheimer:1939ne}.
In GR, solving the TOV equation, we can compute the mass-radius (M-R) relation of the compact star for a given equation of state (EOS), and the observations of the M-R relation can constrain the EOS.
On the other hand, the modified gravity predicts the modified TOV equation and the different M-R relation, allowing us to distinguish the gravitational theories by the astrophysical observations.
Thus, compact star physics is one of the phenomenological applications of the modified gravity theory.

In $f(R)$ gravity theory, the compact star and its M-R relation have been intensively investigated~\cite{Yazadjiev:2014cza, Capozziello:2015yza, AparicioResco:2016xcm, Astashenok:2017dpo, Astashenok:2018iav, Nava-Callejas:2022pip, Feola:2019zqg, Astashenok:2020isy}. 
Among a number of existing studies, we consider models with higher curvature corrections to the Einstein-Hilbert action.
A representative example is the $R^2$ model, which includes the $R^2$ term in addition to the Einstein-Hilbert action, and it is also known as the $R^2$ inflation model~\cite{Starobinsky:1980te}.
It has been suggested that the nonvanishing scalar field surrounds the compact star in the $R^2$ model~\cite{Astashenok:2021xpm, Astashenok:2021peo, Astashenok:2020qds, Astashenok:2021btj, Yazadjiev:2014cza, Astashenok:2017dpo, Astashenok:2018iav, Zhdanov:2024ivr}.
On the other hand, the specific boundary condition near the surface of the compact star predicts the noninteger power-law correction to the Einstein-Hilbert action,
which includes a term $R^{1+\epsilon}$ and $0<\epsilon<1$~\cite{Numajiri:2021nsc}.

The noninteger power (NIP) model has also been studied as the inflationary model~\cite{Artymowski:2014gea};
however, unlike the $R^2$ model, this model possesses the chameleon mechanism.
The chameleon mechanism indicates the scalaron field is very massive in a dense matter environment, and its propagation is screened \cite{Khoury:2003aq, Khoury:2003rn}.
The $R^2$ model does not have the chameleon mechanism and predicts the constant-mass scalaron field.
Although the scalaron field shows exponential decay around the compact star~\cite{Numajiri:2023uif},
this result originates from its massive nature and does not rely on the chameleon mechanism.
Thus, it is intriguing to explore the scalaron field distribution in such higher-curvature corrected models with the chameleon mechanism.
Moreover, it is significant to examine the chameleon mechanism in terms of the thin-shell effect~\cite{Khoury:2003aq, Khoury:2003rn, Brax:2008hh} around the realistic compact object~\cite{Babichev:2009fi, Brax:2017wcj, Kase:2019dqc}.

In this work, we study the scalaron field inside and outside of the compact star in the NIP model.
The chameleon mechanism predicts the thin-shell-like effect, where the scalaron field is expected to decrease sharply near the surface of compact stars.
This behavior of the scalaron field indicates screening the fifth force, which is required to avoid the constraints on the modified theory.
We explore the scalaron behavior near the surface of the compact star and examine the thin-shell effect in the NIP model.
Moreover, we investigate the M-R relation in the NIP model and compare it with those in GR and the $R^2$ model.
We discuss qualitatively the differences in the M-R relation in different gravitational theories for the same EOS.

This paper is organized as follows: 
Sec. II concisely reviews the fundamentals of $f(R)$ gravity theory and introduces the NIP model and the chameleon potential. 
Subsequently, in Sec. III, we discuss the anticipated static spherically symmetric configurations of compact stars and derive the TOV equation, including the necessary boundary conditions. 
Sec. IV details the numerical methods employed in our calculations and presents a comprehensive analysis of the results. 
Finally, in Sec. V, we conclude this paper. 
In this work, we use the $ c=G=1$ unit.


\section{NIP model of \texorpdfstring{$f(R)$}{TEXT} gravity}
\label{sec:basics}  

\subsection{Metric description}

By replacing the Einstein-Hilbert action with the generic function of the scalar curvature $R$, $f(R)$,
the action of the gravitational field with matter field $\Psi$ can be obtained
\begin{align}
\label{eq:action_Jordan}
    S = 
    \frac{1}{2\kappa^2}\int d^{4}x \sqrt{-g}\,f(R)\, + S_m(g_{\mu\nu},\Psi)
    \, ,
\end{align}
where $\kappa^2=8\pi G$ with the gravitational constant $G$.
Variation of Eq.~\eqref{eq:action_Jordan} with respect to the metric $g_{\mu\nu}$ leads to the field equation in $f(R)$ gravity:
\begin{align}
\label{eq:field_equation}
    f_R(R) R_{\mu\nu} -\frac{1}{2} f(R) g_{\mu\nu}
    +(g_{\mu\nu}\square - \nabla_\mu \nabla_\nu)f_R(R) 
    = \kappa^2 T_{\mu\nu}
    \, ,
\end{align}
where $f_R(R)=df/dR$, and $T_{\mu\nu}$ is the energy-momentum tensor defined by varying the action of matter $S_M$
\begin{align}
\label{eq:energy_momentum_tensor}
    T_{\mu\nu}
    =
    -\frac{2}{\sqrt{-g}}\frac{\delta S_M}{\delta g^{\mu\nu}} 
    \, . 
\end{align}
Operating the covariant derivative to Eq.~\eqref{eq:field_equation}, we find the conservation law for $T_{\mu\nu}$,
\begin{align}
\label{eq:conservation_law}
    \nabla^{\mu} T_{\mu\nu} = 0
    \, . 
\end{align}
By multiplying $g^{\mu\nu}$ with both-hand sides of Eq.~\eqref{eq:field_equation}, 
the trace part of the field equation can be read as
\begin{align}
\label{eq:trace_equation}
    \square f_R(R) 
    = 
    \frac{1}{3}\left[ 2f(R)-Rf_R(R)+\kappa^2T\right]
    \, , 
\end{align}
where $T=T^{\mu}_{\mu}$ is the trace of the energy-momentum tensor. 
Note that the curvature $R$ can take nontrivial distribution even in the vacuum ($T_{\mu\nu}=0$).

In this work, we employ a function form for $f(R)$ as follows~\cite{Numajiri:2021nsc}:
\begin{align}
\label{eq:NIP_gravity}
    f(R) = R+a\,R^{1+\frac{1}{b}} 
    \, .
\end{align}
Here, $a$ and $b$ are model parameters, and $b>1$.
The parameter $b$ is dimensionless, while the parameter $a$ has dimension controlled by $b$.
In the unit of length $L$, the dimensions for each quantity are $[R]=L^{-2}$, and $[a]=L^{2/b}$.
This model has a form similar to the $R^2$ model from the viewpoint of higher-order correction of $R$ to GR.
By taking the limit $a\rightarrow 0$ or $b\rightarrow 1$, 
GR or the $R^2$ model are recovered, respectively.
Note that taking the limit $b\rightarrow \infty$,
the NIP model approximates the GR, $f(R) \rightarrow (1+a)R$,
up to the redefinition of the gravitational constant. 

We also comment on the potential origin of this correction \cite{Numajiri:2021nsc}. 
This theory can be seen as the low-energy effective theory of dilatonic gravity defined by the action
\begin{align}
    S_\phi=\int d^4 x \sqrt{-g}\left(\phi R-\frac{\omega(\phi)}{2} \partial_\mu \phi \partial^\mu \phi-V(\phi)\right) ,
\end{align}
with a power-law-type potential 
\begin{align}
    V(\phi) = V_0 \phi^n.
\end{align}
The NIP correction term of the curvature Eq.~\eqref{eq:NIP_gravity} can be obtained by neglecting the kinetic term in the above action and eliminating $\phi$ using the field equation.

Substituting the above NIP model into Eqs.~\eqref{eq:field_equation} and \eqref{eq:trace_equation}, 
the field equation and its trace part are written as
\begin{align}
\label{eq:field_equation_NIP}
\begin{split}
    \kappa^2 T_{\mu\nu}
    &= 
    R_{\mu\nu}-\frac{1}{2}R g_{\mu\nu}
    \\
    & \qquad 
    + a R^{\frac{1}{b}} 
    \left[
        \left(1+\frac{1}{b}\right) R_{\mu\nu} 
        -\frac{1}{2}R g_{\mu\nu}
    \right]
    + a \left(1+\frac{1}{b} \right) (g_{\mu\nu}\square-\nabla_{\mu}\nabla_{\nu})R^{\frac{1}{b}} 
    \, , 
\end{split}
\end{align}
and
\begin{align}
\label{eq:trace_equation_NIP}
    a\left(1+\frac{1}{b}\right)\square R^{\frac{1}{b}} 
    = 
    \frac{1}{3} 
    \left[
        R + a\left(1-\frac{1}{b}\right) R^{1+\frac{1}{b}} +\kappa^2 T
    \right]
    \, . 
\end{align}
One can find in that Eq.~\eqref{eq:trace_equation_NIP},
the scalar curvature $R$ obeys the Klein-Gordon type equation, 
although $R$ is determined by algebraic relation $R = - \kappa^2 T$ in the GR.
This result indicates a scalar DOF in the $f(R)$ gravity theory.
The presence of the scalar DOF in $f(R)$ characterizes the deviations from GR, and as we will see later, the original TOV equation receives significant changes.
We note that in Eq.~\eqref{eq:field_equation_NIP}, we separated the corrections to GR from the Einstein tensor.
The second line in Eq.~\eqref{eq:field_equation_NIP} represents the scalar DOF originating from the gravity sector.


\subsection{Scalar-tensor description in Jordan frame}

To show the scalar DOF explicitly,
we extract the scalar field from the action~\eqref{eq:action_Jordan} and describe $f(R)$ gravity theory in terms of the scalar-tensor theory. 
Defining a new scalar field, called the scalaron, by
\begin{align}
\label{Eq: scalarondef}
    \Phi \equiv f_R(R)
    \, ,
\end{align} 
we find that the action \eqref{eq:action_Jordan} is rewritten as follows:
\begin{align}
\label{eq:action_scalar_field_Jordan} 
    S &= 
    \frac{1}{2\kappa^2}\int d^{4}x \sqrt{-g}\,\left[ \Phi R - Y(\Phi) \right]\, 
    + S_m(g_{\mu\nu},\Psi)
    \, , 
\end{align}
where $Y(\Phi)$ is defined as
\begin{align}
    Y(\Phi) 
    &= 
    \tilde{R}(\Phi)\Phi - f(\tilde{R}(\Phi))
    \, .
\end{align}
As in Eq.~\eqref{eq:action_scalar_field_Jordan}, 
the $f(R)$ gravity theory can be regarded as GR nonminimally coupling the scalaron field.
It is now apparent that in addition to two DOFs from the Einstein-Hilbert action, 
there is an additional DOF originating from the scalaron field~\cite{Numajiri:2023uif}.

The field equation with respect to metric can be obtained 
by varying the action~\eqref{eq:action_scalar_field_Jordan} with respect to $g_{\mu\nu}$:
\begin{align}
\label{eq:equation_motion1}
    R_{\mu\nu}-\frac{1}{2}g_{\mu\nu}R
    = 
    \frac{\kappa^2}{\Phi}(T_{\mu\nu}+T^{(\Phi)}_{\mu\nu})
    \, .
\end{align}
$T^{(\Phi)}_{\mu\nu}$ is an effective energy-momentum tensor of the scalar field, 
\begin{align}
\label{eq:energy-momentum tensor for scalar}
    T^{(\Phi)}_{\mu\nu}
    =
    \frac{1}{\kappa^2}
    \left[
        \nabla_{\mu}\nabla_{\nu}\Phi -g_{\mu\nu}\square\Phi - \frac{1}{2}g_{\mu\nu} Y(\Phi)
    \right]
    \, .
\end{align}
Moreover, the variation of the action~\eqref{eq:action_scalar_field_Jordan} with respect to $\Phi$ leads to $R = Y_{\Phi} (\Phi)$, where $Y_{\Phi} = \partial Y/ \partial \Phi$.
Combining it with the trace of Eq.~\eqref{eq:equation_motion1}, we obtain the field equation with respect to the scalaron field:
\begin{align}
\label{eq:equation_motion2}
    \square \Phi 
    = 
    \frac{d V_{\mathrm{eff}}}{d \Phi}
    \, .
\end{align}
where we define the effective potential $V_{\mathrm{eff}}(\Phi, T)$ of the scalar field:
\begin{align}
\label{eq:definition_eff_potential}
    \frac{d V_{\mathrm{eff}}(\Phi, T)}{d \Phi} 
    \equiv 
    \frac{1}{3}
    \left[\Phi R(\Phi) - 2Y(\Phi)+\kappa^2 T \right]
    \, .
\end{align}
We note that $V_{\mathrm{eff}}$ explicitly depends on $T$, and we denote it by $V_{\mathrm{eff}}(\Phi, T)$.

Finally, we consider the effective mass of the scalaron field.
The mass is related to the second derivative potential given by
\begin{align}
\label{eq:definition_eff_mass}
\begin{split}
    \frac{d^2 V_{\mathrm{eff}}}{d \Phi^2} 
    & \equiv 
    \frac{1}{3} \frac{d}{d\Phi}
    \left[\Phi R(\Phi) - 2Y(\Phi)+\kappa^2 T \right]
    \\
    & =
    \frac{1}{3} 
    \left[
        \frac{F_R (R)}{F_{R R} (R)} - R
    \right]
    \, .
\end{split}
\end{align}
By convention, the second derivative at the potential minimum is used to define the effective mass,
\begin{align}
\label{eq:definition_eff_mass1}
    m_{\Phi}^2 
    &\equiv 
    \left. \frac{d^{2} V_{\mathrm{eff}}}{d \Phi^2} \right|_{{\Phi}=\Phi_{\min}}
    \, ,
\end{align}
where the minimum of the effective potential can be calculated by using the stationary condition,
\begin{align}
\label{eq:stationary_condition}
    \left. \frac{dV_{\mathrm{eff}}}{d\Phi}=0 \right|_{{\Phi}=\Phi_{\min}}
    \, .
\end{align}
Note that for a given $T$, we can search for $\Phi=\Phi_{\min}$ at the potential minimum.

\begin{figure}[htbp]
\begin{tabular}{cc}
    \begin{minipage}{0.5\linewidth}
        \centering
        \includegraphics[keepaspectratio, width=0.9\linewidth]{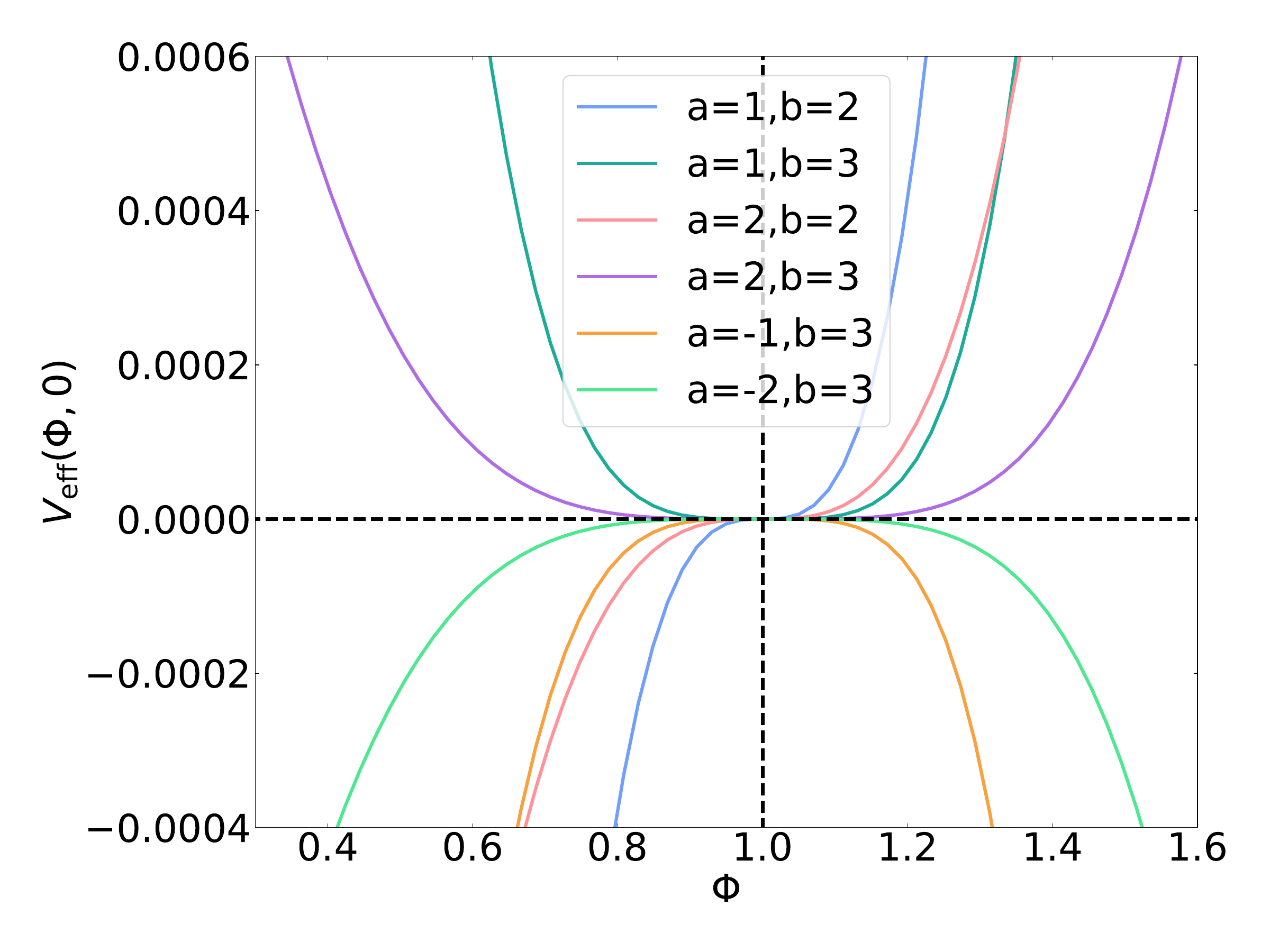}%
        \subcaption{
            Varying $a$ and $b$ with $T=0$.
        }
        \label{fig:chame_pot_JO_ab}
    \end{minipage}
    &
    \begin{minipage}{0.5\linewidth}
    \centering
        \includegraphics[keepaspectratio, width=0.9\linewidth]{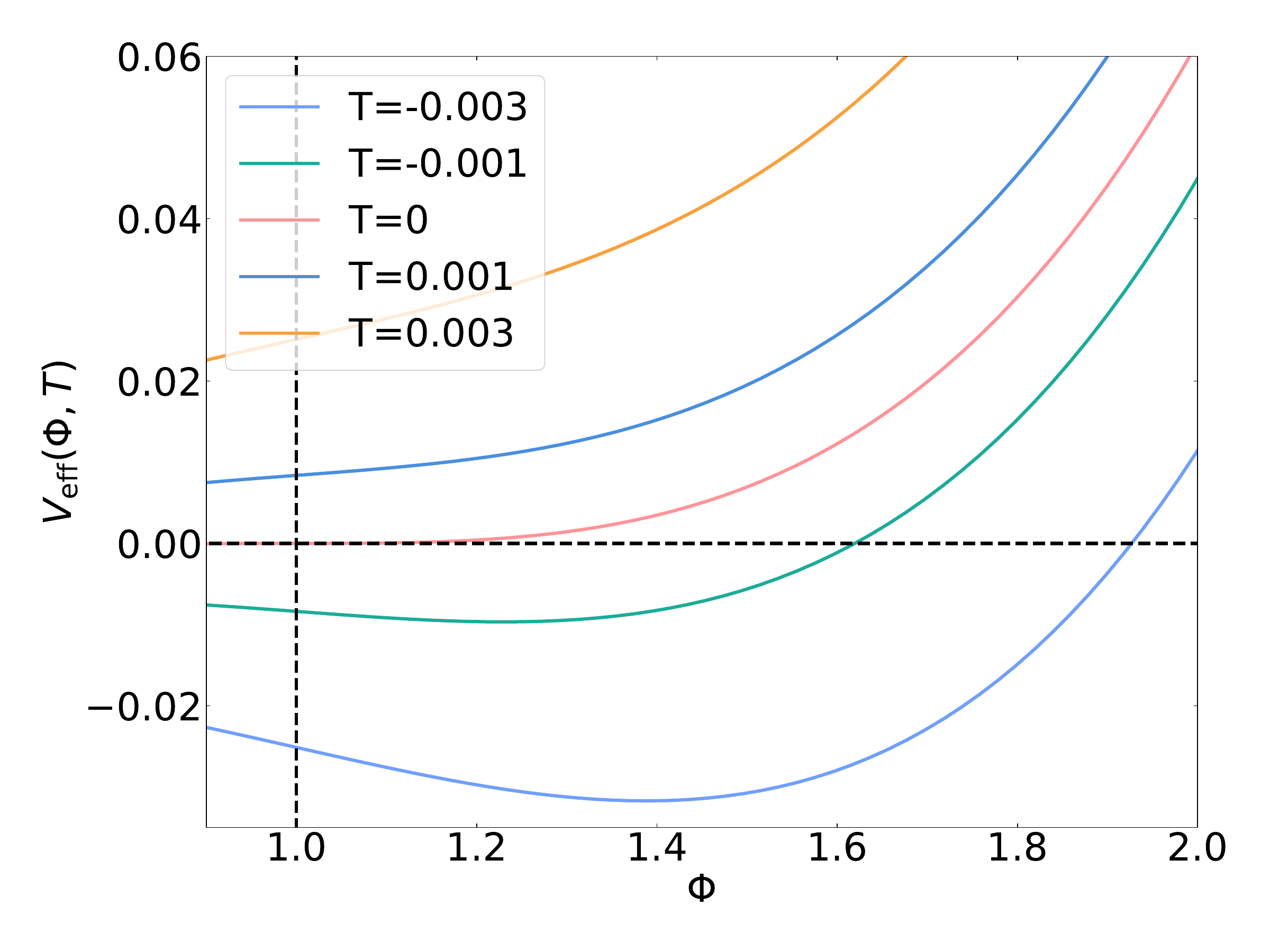}%
        \subcaption{
            Varying $T$ with $a=1r_{g},b=2$.
        }
    \label{fig:chame_pot_JO_T}
    \end{minipage}
\end{tabular}
\caption{
    The effective potential~\eqref{eq:chameleon_potential_in_NIP} in the NIP model of $f(R)$ gravity, 
    normalized by the Schwarzschild radius of the solar mass $r_g = GM_{\odot}$.  
    (a) The potential shape becomes steeper as $|a|$ or $b$ decreases in the vacuum region, 
    and whether $b$ is odd or even controls the symmetry of the potential.
    (b) The minimum of the potential changes according to $T$.
}
\label{fig:chame_pot_JO}
\end{figure}

We consider the field equation of the scalaron and observe the chameleon mechanism \cite{Khoury:2003aq, Khoury:2003rn} in the NIP model.
By substituting Eq.~\eqref{eq:NIP_gravity} into Eq.~\eqref{Eq: scalarondef},
the scalaron field is given as
\begin{align}
\label{eq:scalaron_def}
    \Phi=1+a\left( 1+\frac{1}{b} \right)R^{\frac{1}{b}}
    \, . 
\end{align}
Here, the right-hand side of Eq.~\eqref{eq:equation_motion1} suggests that $G/\Phi = G / f_{R}$ gives the effective gravitational constant.
To avoid the antigravity (repulsive gravitational interaction), we impose $f_R>0$.
This condition indicates that the scalar curvature $R$ should satisfy the following condition:
\begin{align}
\label{eq:non_antigravity_condition}
    \left\{ \ 
    \begin{aligned}
        0\leq\, &R\,\leq \left(\frac{b}{\left\lvert a\right\rvert (1+b)}\right)^b \quad  
        & \text{for} \ a<0
        \, , \\ 
        & 0 \leq R \quad 
        & \text{for} \ a>0  
        \, . 
    \end{aligned}
    \right.
\end{align}
We obtain the field equation for the scalaron field~\eqref{eq:equation_motion2} in the NIP model, 
\begin{align}
\label{eq:equation_motion_NIP}
    \square\Phi
    &=
    \frac{1}{3}\left\{\frac{(b-1)\Phi+2}{b+1}
    \left[\frac{b(\Phi-1)}{a(b+1)}\right]^{b}+\kappa^{2}T\right\}
    \, ,
\end{align}
and the effective potential $V_{\mathrm{eff}}$~\eqref{eq:definition_eff_potential} is given as
\begin{align}
\label{eq:chameleon_potential_in_NIP}
    V_{\mathrm{eff}}(\Phi,T) 
    = 
    \frac{1}{3}
    \left\{
        \frac{\left[(b-1)\Phi+3\right](\Phi-1)}{(b+1)(b+2)} 
        \left[\frac{b(\Phi-1)}{a(b+1)}\right]^b + \kappa^2T\Phi
    \right\}
    \, .
\end{align}
We plot the effective potential in \figref{fig:chame_pot_JO}.

The effective mass can be defined at the potential minimum, 
\begin{align}
\label{eq:chameleon_Mass_NIP}
\begin{split}
    m_{\Phi}^2
    &=
    \frac{b[1+(b-1)\Phi_{\min}]}{3a(b+1)}\left[\frac{b(\Phi_{\min}-1)}{a(b+1)}\right]^{b-1}
    \, .
\end{split}
\end{align}
$\Phi_{\min}$ is determined by $dV_{\mathrm{eff}}/d\Phi=0$, and it depends on $T$.
As shown in \figref{fig:chame_pot_JO_T},
the curvature of the effective potential changes depending on the value of $T$.
We find that $\Phi_{\min}>1$ and $m_{\Phi}^2>0$ for $T<0$,
which illustrate the so-called chameleon mechanism~\cite{Khoury:2003aq,Khoury:2003rn}.
It is worth mentioning that
there is only one solution $\Phi_{\min}=1$ to the stationary condition in the vacuum $T=0$,
where the scalaron behaves as a massless particle $m_{\Phi}=0$.

Note that in the following calculation, we do not introduce the frame transformation of the metric.
That is, we work in the Jordan frame, 
where the gravitational action is defined as Eq.~\eqref{eq:action_scalar_field_Jordan},
while we discuss the NIP model in the Einstein frame in the Appendix.
Moreover, Eq.~\eqref{eq:chameleon_Mass_NIP} implies that the scalaron is tachyonic or ill defined for $a<0$ and general $b$.
Thus, we focus on the region $a>0$,
where Eq.~\eqref{eq:scalaron_def} leads to the scalaron-field value $\Phi\geq 1$ and the scalar curvature $R \geq 0$.


\section{Static and spherically symmetric star under NIP gravity}

\subsection{The configuration of system}

We consider a static and spherically symmetric star composed of the perfect fluid.
The geometry inside and outside the star is characterized by the following spacetime metric $g_{\mu\nu}$:
\begin{align}
\label{eq:Metric_sss}
    ds^2 
    = 
    g_{\mu\nu}dx^{\mu}dx^{\nu}
    = 
    -e^{2\nu(r)}dt^2+e^{2\lambda(r)}dr^2+r^2d\Omega^2
    \, ,
\end{align}
where $x^{\mu}=\{t,r,\theta,\varphi\}$,
and $d\Omega^2=d\theta^2+\sin^2\theta d\varphi^2$ is the line element of the ywo-sphere.
The $(t,t)$ and $(r,r)$ components of the metric, $\nu(r)$ and $\lambda(r)$, depend only on the radial coordinate $r$.

The conserved energy-momentum tensor inside the star is defined as 
\begin{align}
\label{eq:mass_Energy-momentum_tensor}
    T_{\mu\nu} = [\epsilon(r)+p(r)]u_{\mu}u_{\nu}+p(r)g_{\mu\nu}
    \, , 
\end{align}
where $\epsilon(r)$ and $p(r)$ are the total energy density and the pressure, respectively.
$u^{\mu} = \{e^{-\nu},0,0,0\}$ is the four velocity of static fluid. 
The radius of the star $r=r_s$ is defined as the radial coordinate where the pressure vanishes $p(r_s)=0$. 
The matter composing compact stars is assumed to be the perfect fluid, 
and it is characterized by the EOS $p=p(\rho)$ that describes the physical properties of the matter.
In this work, we use the (piecewise) polytrope EOS~\cite{Read:2008iy}:
\begin{align}
\label{eq:Piecewise_Polytropic_EOS}
    p(\rho)=K \rho^\Gamma
    \, .
\end{align}
Here $a$, $K$, and $\Gamma$ are parameters characterizing the EOS.
The total energy density $\epsilon(r)$ and rest-mass density $\rho(r)$ are related through the first law of thermodynamics,
and for polytrope EOS, we find the following relation
\begin{align}
\label{eq:relation_eps_rho}
    \epsilon(\rho)=(1+a)\rho + \frac{K}{\Gamma-1} \rho^{\Gamma}
    \, .
\end{align}
When implementing the numerical calculation with the piecewise polytope EOS, 
we define the discrete intervals and assign the index $i$ for $a$, $K$, and $\Gamma$ in each interval.

We impose the junction condition at the surface of the compact star $r=r_s$. 
For the matter sector, we assume 
\begin{align}
    \left[T_{\mu\nu}\right] = 0
    \, ,
\end{align}
where $\qty[\dots]$ denotes the jump over some timelike hypersurface. 
This condition is also consistent with our assumption of the polytope EOS, and we do not study the thin-shell collapse with a delta-function-like discontinuity.
For the gravity sector, 
there are many existing works~\cite{Olmo:2020fri,Feng:2017hje,Senovilla:2013vra} that have discussed the junction condition in $f(R)$ gravity.  
The first and second fundamental forms across a timelike hypersurface are the same as the usual junction condition in GR~\cite{Ganguly:2013taa}
\begin{align}
    \left[g_{\mu\nu}\right]=0\, ,\quad \left[K_{\mu\nu}\right] = 0\, ,
\end{align}
where $K_{\mu\nu}$ is extrinsic curvature.
If this condition is violated, then the spacetime is singular at this surface. 
Furthermore, the field equation Eqs.~\eqref{eq:trace_equation} or \eqref{eq:equation_motion2} of additional scalar DOF demands the curvature continuity conditions \cite{Deruelle:2007pt}
\begin{align}
    \left[R\right] = 0\, ,\quad \left[\nabla_{\mu} R\right]=0\, .
\end{align}
These conditions indicate no discontinuity at the surface for the metric, matter field, and curvature. Because the numerical solution of inner and outer are calculated independently, these conditions should
be imposed when they are connected.


\subsection{Modified TOV equations}

We derive the modified TOV equation in the NIP model.
First of all,
$r$ component of the conservation law as in Eq.~\eqref{eq:conservation_law} leads to
\begin{align}
\label{eq:continuity_NIP} 
    0 = p^{\prime}+\nu^{\prime} \left(\epsilon + p \right)
    \, ,
\end{align}
which is independent of model of $f(R)$ gravity.
Substituting the spacetime metric $g_{\mu\nu}$ in Eq.~\eqref{eq:Metric_sss} 
into Eqs.~\eqref{eq:field_equation_NIP} and \eqref{eq:trace_equation_NIP},
we obtain the differential equations for $\lambda(r), \nu(r)$, and $R(r)$:  
\begin{align}
\label{eq:lambda'_NIP}
\begin{split}
    \lambda^{\prime}
    &= \frac{1}{4 b^2 r R^2 \left[a (b+1) R^{\frac{1}{b}}+b \right] +2 a b (b+1) r^2
    R^{\frac{1}{b}+1} R^{\prime}} 
    \\
    &\quad
    \left[ e^{2 \lambda }
        \left(
            a b^2 r^2 R^{\frac{1}{b}+3}-2 b^2 R^2 \left(a (b+1)
            R^{\frac{1}{b}}-b \kappa ^2 r^2 \epsilon +b\right)
        \right)
        +2 b^2 R^2
        \left(a (b+1) R^{\frac{1}{b}}+b\right)
    \right. 
    \\
    &\hspace{47mm}
    \left.
        +2 a b (b+1) r R^{\frac{1}{b}+1} \left(2 R'+r R^{\prime\prime}\right)
        -2 a \left(b^2-1\right) r^2 R^{\frac{1}{b}} R^{\prime 2}
    \right]
    \, , 
\end{split}
\end{align}
\begin{align}
\label{eq:nu'_NIP}
\begin{split}
    &\nu^{\prime}
    = \frac{1}{4 b r R \left[a (b+1) R^{\frac{1}{b}}+b\right]+2 a (b+1) r^2 R^{\frac{1}{b}} R^{\prime}}
    \\
    &\qquad 
    \left[ e^{2 \lambda } 
        \left(
            -a b r^2 R^{\frac{1}{b}+2}
            +2 b R \left(a (b+1) R^{\frac{1}{b}}+b \kappa ^2
            r^2 p+b\right)                              
        \right)
        -2 b R \left(a (b+1) R^{\frac{1}{b}}+b\right)
    \right.
    \\
    &\hspace{115mm}
    \left.
        -4 a (b+1) r R^{\frac{1}{b}} R^{\prime}
    \right]
    \, , 
\end{split}
\end{align}
\begin{align}
\label{eq:dynamicalR_NIP}
\begin{split}
    &R^{\prime\prime}
    = \frac{(b-1) R^{\prime 2}}{b R}
    +\left(\lambda^{\prime}+\frac{1}{r}\right) R^{\prime}
    \\
    &\hspace{35mm}
    +\frac{bR^{1-\frac{1}{b}}}{2a(1+b)r^2}
    (b+a(1+b)R^{\frac{1}{b}}) 
    [e^{2 \lambda } \left(r^2 R-4\right)-2 r \lambda^{\prime}
    +6 r \nu^{\prime}+4]
    \, .
\end{split}
\end{align}
Here, we denote the derivative with respect to $r$ by the prime.
The second derivative of $\nu$ in field equations has been deleted by using the following relation
\begin{align}
    R=
    e^{-2\lambda}
    \left[
        -2\nu^{\prime\prime}-2\nu^{\prime}(\nu^{\prime}-\lambda^{\prime})
        -\frac{4(\nu^{\prime}-\lambda^{\prime})}{r}+\frac{2e^{2\lambda}-2}{r^2}
    \right]
    \, .
\end{align}

As displayed above, we have five functions, $\lambda(r)$, $\nu(r)$, $R(r)$, $\epsilon(r)$, and $p(r)$ to be solved 
and four equations~\eqref{eq:continuity_NIP} -- \eqref{eq:dynamicalR_NIP}.
To complete the system, we impose a specific EOS capable of determining the relation between the total energy density $\epsilon(r)$ and the pressure $p(r)$.
Thus, we need to solve four ordinary differential equations (ODEs) for four functions $\lambda(r)$, $\nu(r)$, $R(r)$, and $p(r)$ with the assumed EOS.


\subsection{Asymptotic solution and scalar hair}

To consider the boundary condition outside the star, we analyze the asymptotic behavior of the spacetime.
We begin with a realistic assumption that the spacetime geometry shows asymptotic flatness, where the scalar curvature $R$ is expected to go to zero as $r \rightarrow \infty$.
Based on this assumption, Eq.~\eqref{eq:scalaron_def} suggests the scalaron field $\Phi$ goes to unity in vacuum.
This behavior is actually consistent with the physical insight that the scalaron field rolls down the effective potential and reaches $\Phi=1$,
which also satisfies the stationary condition in the vacuum.

The scalaron characterizes the deviation from GR, and $\Phi=f_{R}=1$ corresponds to the GR limit.
Thus, the NIP model approaches GR far from the compact star,
as the scalaron field approaches $\Phi=1$.
Although the $f(R)$ gravity does not possess Birkhoff-Jebsen's theorem in general,
the NIP model asymptotically shows it.
Similar results can also be found in the $R^2$ model~\cite{Numajiri:2023uif}.
Then, the spacetime solution is the asymptotically Schwarzschild one at $r \rightarrow \infty$: 
\begin{align}
    \label{eq:sch_solution0}
    &\lim_{r\rightarrow\infty} e^{2\nu(r)}
    =
    \lim_{r\rightarrow\infty}e^{-2\lambda(r)}
    = 1-\frac{2M}{r} 
    \, , \nn
    &\lim_{r\rightarrow\infty} R(r)
    =
    \lim_{r\rightarrow\infty}R^{\prime}(r)
    =0 
    \, .
\end{align}
The constant $M$ corresponds to the Schwarzschild mass. 
We can also define a mass within radius $r$ as
\begin{align}
\label{eq:Sch_mass}
    m(r)\equiv \frac{r}{2}\left(1-e^{-2\lambda(r)}\right).
\end{align}
where $m(\infty)$ equals to the Schwarzschild mass $M$. 
It should be noted that in the $f(R)$ gravity $m(r_s)$ does not provide the proper stellar mass as an observable since it does not include the contribution of the scalaron (or the curvature), which stems from the absence of Birkhoff-Jebsen's theorem.

Although the one-way integration method to be adopted does not need to identify the form of the asymptotic solution, 
we briefly discuss the asymptotic behavior of the scalaron field $\Phi$ outside the compact star.
Assuming the asymptotic Minkowski spacetime and a vacuum $T=0$, far enough away from the compact star,
we find Eq.~\eqref{eq:equation_motion_NIP} reduces into the following form:
\begin{align}
    \Box (\Phi-1)
    &=
    \frac{(b-1)}{3(b+1)}\left[\frac{b}{a(b+1)}\right]^{b} (\Phi-1)^{b+1} 
    +\frac{1}{3}\left[\frac{b}{a(b+1)}\right]^{b} (\Phi-1)^{b}
    \, .
\end{align}
We expect $\Phi\approx1$ away from the compact star,
and when $0<\Phi-1\ll1$ and $b>1$, we find $(\Phi-1)^{b+1} \ll (\Phi-1)^{b}$.
Thus, the scalaron field equation around $\Phi=1$ at the leading order is given as
\begin{align}
\label{eq:scalaron_asympt}
    (\Phi-1)^{\prime \prime}
    + \frac{2}{r}  (\Phi-1)^{\prime}
    &=
    \frac{1}{3}\left[\frac{b}{a(b+1)}\right]^{b} (\Phi-1)^{b}
    \, .
\end{align}
The above equation determines the scalaron behavior around $\Phi \approx 1$.

When $a$ is very large $a \gg 1$, the right-hand side of Eq.~\eqref{eq:scalaron_asympt} is negligible.
In this case, Eq.~\eqref{eq:scalaron_asympt} approximates the Klein-Gordon equation for a free particle, and the scalaron field behaves like the Coulomb potential, $\Phi (r) \sim C_1/r + C_2$, where $C_1$ and $C_2$ are integration constants.
When $a$ is very small $a \ll 1$, 
the right-hand side of Eq.~\eqref{eq:scalaron_asympt} is dominant.
Then, the scalaron may rapidly go to the potential minimum in vacuum $\Phi(r) \rightarrow 1$, corresponding to the GR limit.
We can expect that the scalaron field $\Phi(r)$ is a decreasing function with respect to $r$ and may decay more quickly for the smaller $a$.


\subsection{Boundary conditions}

We solve the modified TOV equations as a multiboundary value problem and specify boundary conditions at $r=0$ and $r=\infty$.
We have four differential equations with an EOS, and three of them are the first-order ODEs~\eqref{eq:continuity_NIP},~\eqref{eq:lambda'_NIP}, and ~\eqref{eq:nu'_NIP} for $p(r)$, $\lambda(r)$, and $\nu(r)$. 
One of them is a second-order ODE for the curvature $R(r)$.
Thus, we need to specify five boundary conditions.
To avoid the conical singularity at the center of star $r=0$, we demand that 
\begin{align}
    \lambda(0)=0 
    \, , \quad 
    R^{\prime}(0)=0 
    \, .
\end{align} 
We still have three undetermined values at the center of the star, 
\begin{align}
    \nu(0)=\nu_{c} 
    \, , \quad 
    R(0)=R_{c}
    \, , \quad
    p(0)=p_{c}
    \, .
\end{align}
Equation~\eqref{eq:sch_solution0} determines $\nu_c$ and $R_c$ so that $\nu(r)$ and $R(r)$ approach the Schwarzschild solution at the radial infinity. 
Note that the EOS gives the central pressure $p_c$ when the central rest-mass density $\rho_c$ is specified. 
Thus, $R_c$ and $\nu_c$ are determined by the boundary conditions at radial infinity, and we treat them as the shooting parameters.

To solve the modified TOV equation smoothly, 
we need to impose the asymptotic behaviors around the center.
By expanding the solution of the Eqs.~\eqref{eq:continuity_NIP}--\eqref{eq:dynamicalR_NIP} around $r=0$, 
four functions behave as 
\begin{align}
    \begin{aligned}
        &\lambda (r)\simeq\lambda_0+\lambda_1r+\frac{1}{2}\lambda_2r^2\,, &\nu(r)\simeq\nu_0+\nu_1r+\frac{1}{2}\nu_2r^2\, , \\
        &R(r)\simeq R_0 +R_1r+\frac{1}{2}R_2r^2\, , &p(r)\simeq p_0+p_1r+\frac{1}{2}p_2r^2\, .
    \end{aligned}
\end{align}
Substituting them into Eqs.~\eqref{eq:lambda'_NIP}--\eqref{eq:dynamicalR_NIP} determines the each coefficient:
\begin{align}
\begin{split}
    \lambda_0 = 0
    \, , \quad 
    \nu_0 = \nu_c
    \, ,\quad 
    R_0 = R_c
    \, ,\quad 
    p_0 = p_c
    \, ,\quad 
    \epsilon_0 = \epsilon_c 
    \, , 
\end{split}
\\
\begin{split}
    \lambda_1=\nu_1=R_1 = p_1=0 
    \, , 
\end{split}
\\
\begin{split}
    \lambda_2
    &= 
    \frac{2bR_0+4b\kappa^2\epsilon_0+6b\kappa^2p_0+a(2b+1)R_0^{\frac{1}{b}+1}}
    {18\left[a(b+1)R_0^{\frac{1}{b}}+b\right]}
    \, ,\\ 
    \nu_2 
    &= 
    -\frac{bR_0-4b\kappa^2\epsilon_0-6b\kappa^2p_0+a(b+2)R_0^{\frac{1}{b}+1}}
    {18\left[a(b+1)R_0^{\frac{1}{b}}+b\right]}
    \, ,\\ 
    R_2 
    &= 
    \frac{\left[a(b-1)R_0^{\frac{1}{b}+1}-b\kappa^2\epsilon_0+3b\kappa^2p_0+bR_0\right]bR_0^{\frac{b-1}{b}}}
    {9a(b+1)}
    \, ,\\
    p_2 
    &= \frac{(p_0+\epsilon_0)\left[bR_0-4b\kappa^2\epsilon_0-6b\kappa^2p_0+a(b+2)R_0^{\frac{1}{b}+1}\right]}
    {18\left[a(b+1)R_0^{\frac{1}{b}}+b\right]}
    \, .
\end{split}
\end{align}

Regarding boundary conditions at $r=\infty$, we rely on Eq.~\eqref{eq:sch_solution0} to determine $\nu_{c}$ and $R_{c}$.
In summary, we determine the five boundary conditions given by 
\begin{align}
\label{eq:boundary_conditions}
\begin{split}
    &\lambda(0) = 0 \, , \quad p(0) = p_c \, , \quad R^{\prime}(0) = 0 \, , \\ 
    &\nu(\infty)= \frac{1}{2} \ln \left(1 - \frac{2M}{r}\right) \, , \quad R(\infty) = 0 \, ,
\end{split}
\end{align} 
and $p_c$ is given by hand. 
We use the general shooting method; from the origin $r=0$, we start the shooting until the solution satisfies the two boundary conditions in the second line of Eq.~\eqref{eq:boundary_conditions}.


\subsection{Numerical implementations}

To numerically solve the modified TOV equations, we consider the radial direction divided into three regions.
The first one is inside the star, $r\in\left[0,r_s\right]$, where $r_s$ is the surface of the star.
For outside the star,  $r\in\left(r_s,\infty\right)$,
we define the radius $r_\Phi$ where the scalaron field is small enough to be neglected. 
Thus, we have $\left[0,r_s\right]$, $\left(r_s,r_\Phi\right]$, and $\left(r_\phi,\infty\right)$.
The origin point in the shooting method is set as $r_c=10^{-3}r_g$.

In the above configuration, 
we solve Eqs.~\eqref{eq:continuity_NIP} -- \eqref{eq:dynamicalR_NIP} from $r_c$ to $r_\Phi$.
We use the fourth-order Runge-Kutta method supported by the package $SciPy$ in P\textsc{ython} to solve the ODE system.
First, we define the dimensionless variables by the following characteristic scales:
\begin{align}
\begin{split}
    &M_g = M_{\odot} \simeq 1.99 \times 10^{33}~\mathrm{g}
    \, , \\
    &r_g=\frac{GM_{\odot}}{c^2}\simeq1.48 \times 10^5~\mathrm{cm}
    \, , \\
    &\rho_g = \frac{M_g}{r_g^3}\simeq 6.18\times 10^{17}~\mathrm{g\,cm^{-3}}
    \, , \\ 
    &p_g = \frac{M_g c^2}{r_g^3}\simeq 5.55\times 10^{38}~\mathrm{g\,cm^{-1}s^{-2}}
    \, , \\
    &R_g = \frac{\rho_g G}{c^2}\simeq 4.59\times 10^{-11}~\mathrm{cm^{-2}}
    \, , \\ 
    &m_{\Phi g}=R_g^{\frac{1}{2}}\simeq 0.34~\mathrm{eV}
    \, .
\end{split}
\end{align}

For the center $r = r_c$, 
we constantly change the value of $R_c$ with given the $\nu_c$ and $p_c$ to solve the ODE system 
until it reaches the condition~\eqref{eq:boundary_conditions} and take a rescaling for $\nu$ to satisfy the Schwarzschild solution.
In addition, for the EOS $p=p(\rho)$, we used the piecewise form~\cite{Read:2008iy} of SLy EOS~\cite{Alford:2004pf}, 
which corresponds to the neutron star with a quark matter core with quantum chromodynamics correction. 
There is the upper value for $p_c$, which EOS itself constrains that the sound speed $v^2 = \frac{dp}{d\epsilon}$ does not exceed the light speed $c^2$. 
The positive chameleon mass in the Einstein frame demands a boundary of central pressure $p_c$.
For the value of the parameter $a$ and $b$, 
we take $a= 0, 0.5r_g^{\frac{2}{b}}, 1r_g^{\frac{2}{b}}, 10r_g^{\frac{2}{b}}, 100r_g^{\frac{2}{b}}$ and $b = 2$. 
Note that we choose the parameter $\alpha$ to be on the astrophysical scale.
One could derive limits on the parameter from cosmological observations by assuming higher energy scales and adapting the higher-order curvature term to inflation. 
However, we consider the lower energy scales, especially the compact star scale, due in part to numerical difficulties.


\section{Numerical Results and Discussions}
\label{sec:results}

\subsection{Spacetime geometry}

\begin{figure}[htbp]
\begin{tabular}{cc}
    \begin{minipage}{0.5\linewidth}
        \centering
        \includegraphics[keepaspectratio, width=0.9\linewidth]{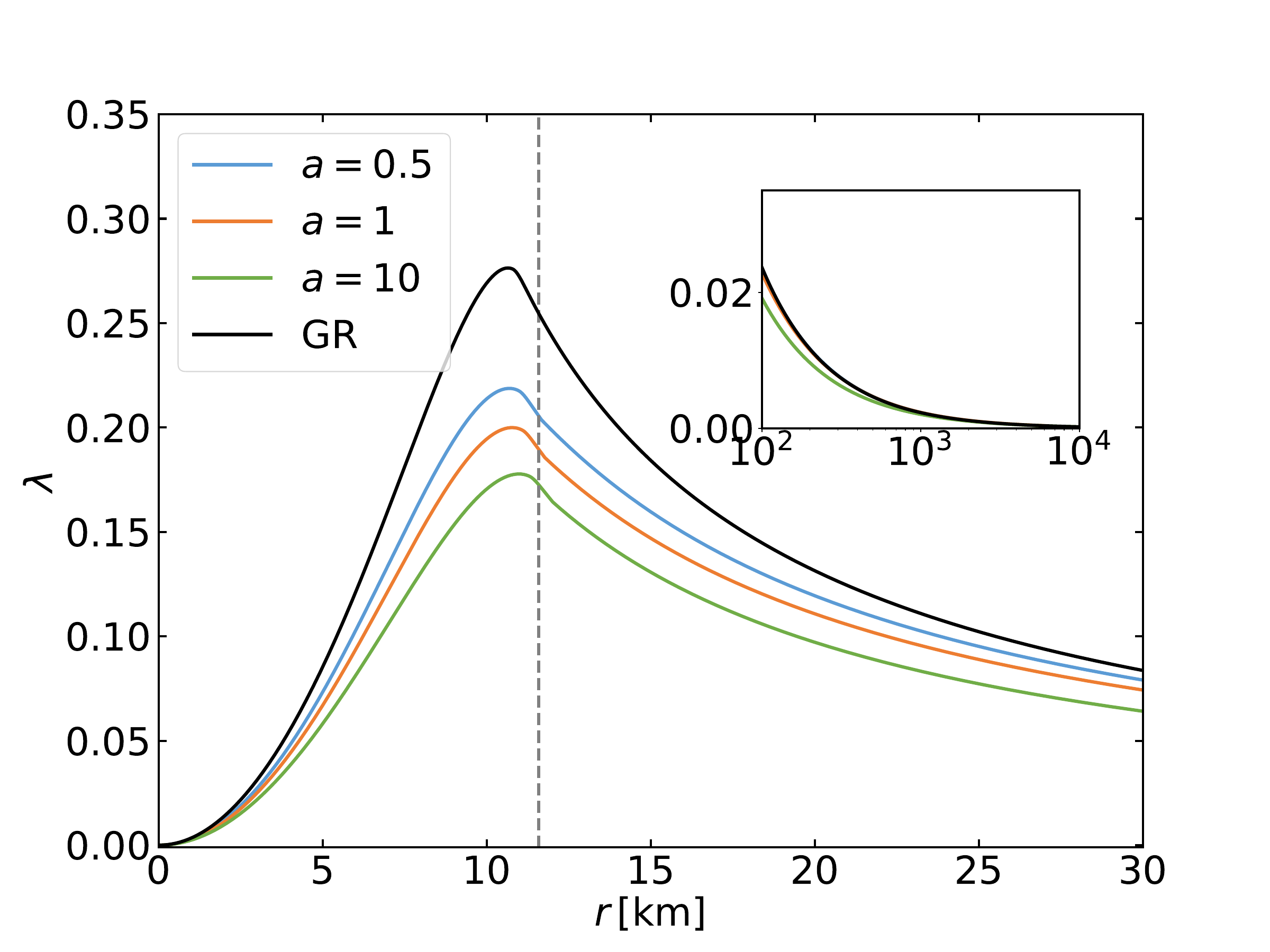}%
        \subcaption{
            Metric component $\lambda(r)$
        }
        \label{fig:lambda_r_15}
    \end{minipage}
    &
    \begin{minipage}{0.5\linewidth}
        \centering
        \includegraphics[keepaspectratio, width=0.9\linewidth]{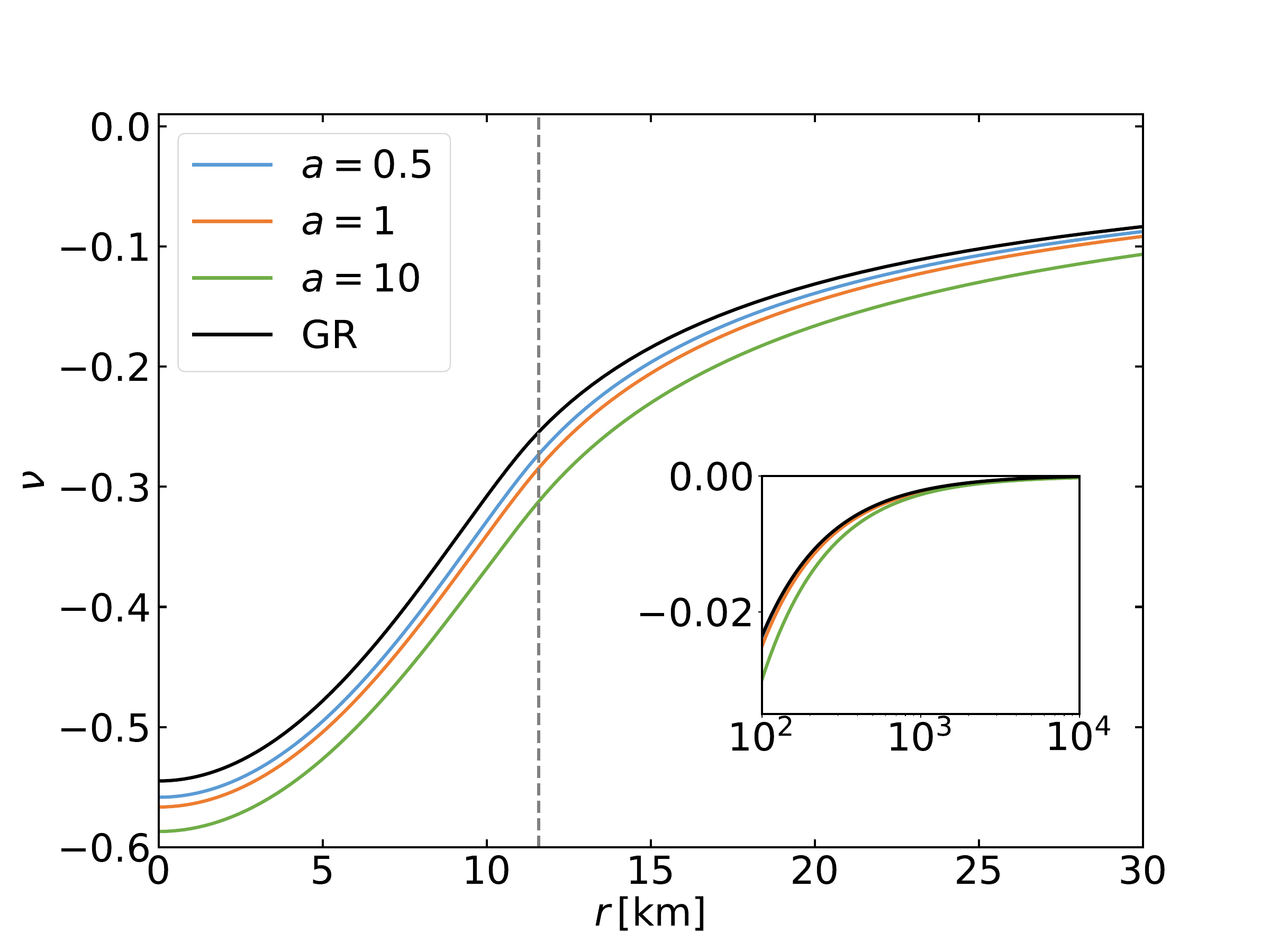}%
        \subcaption{
            Metric component $\nu(r)$
        }
        \label{fig:nu_r_15}
    \end{minipage}
    \\
    \begin{minipage}{0.5\linewidth}
        \centering
        \includegraphics[keepaspectratio, width=0.9\linewidth]{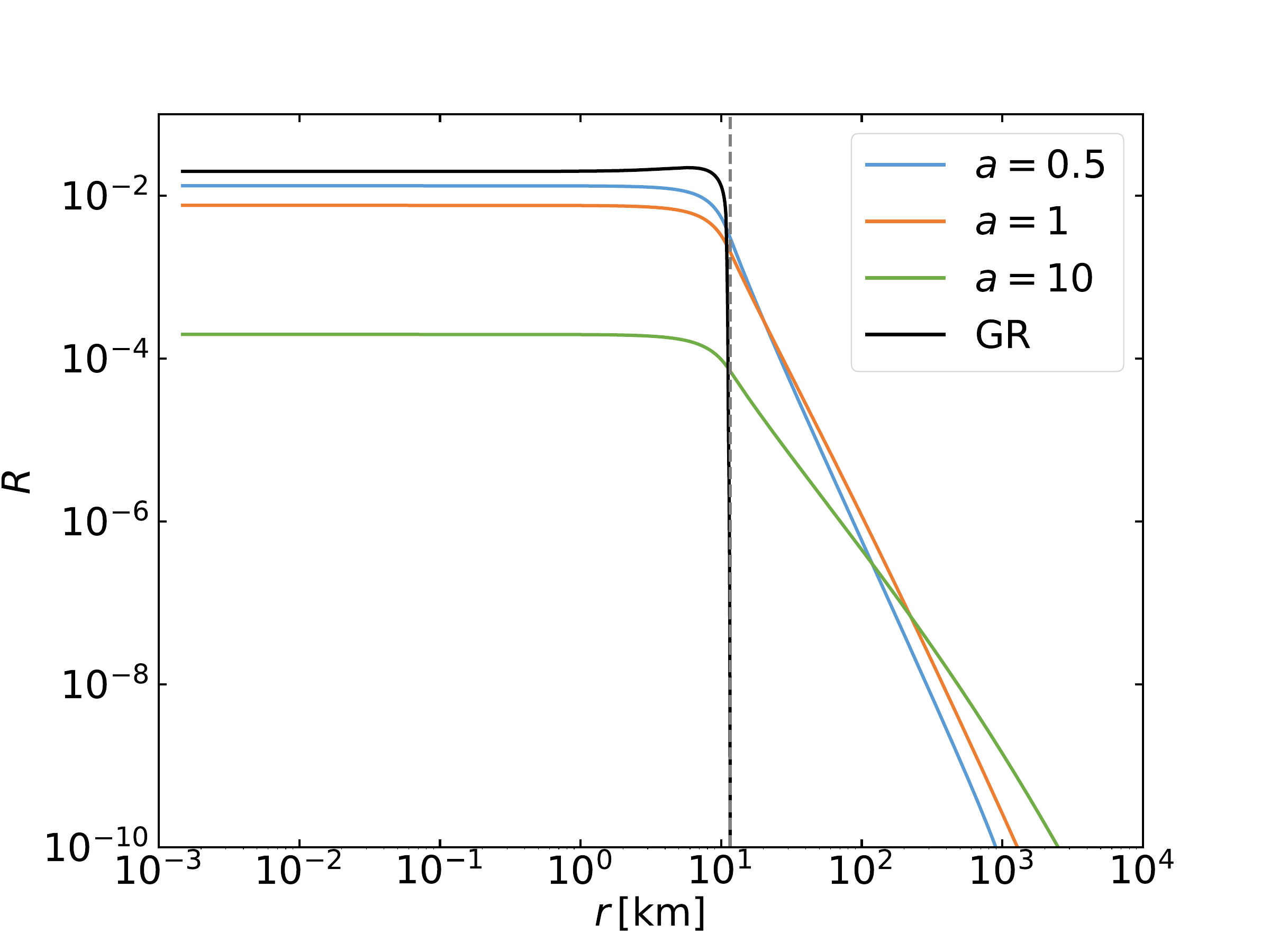}%
        \subcaption{
            Curvature profile $R(r)$
        }
        \label{fig:R_r_15}
    \end{minipage}
    &
    \begin{minipage}{0.5\linewidth}
        \centering
        \includegraphics[keepaspectratio, width=0.9\linewidth]{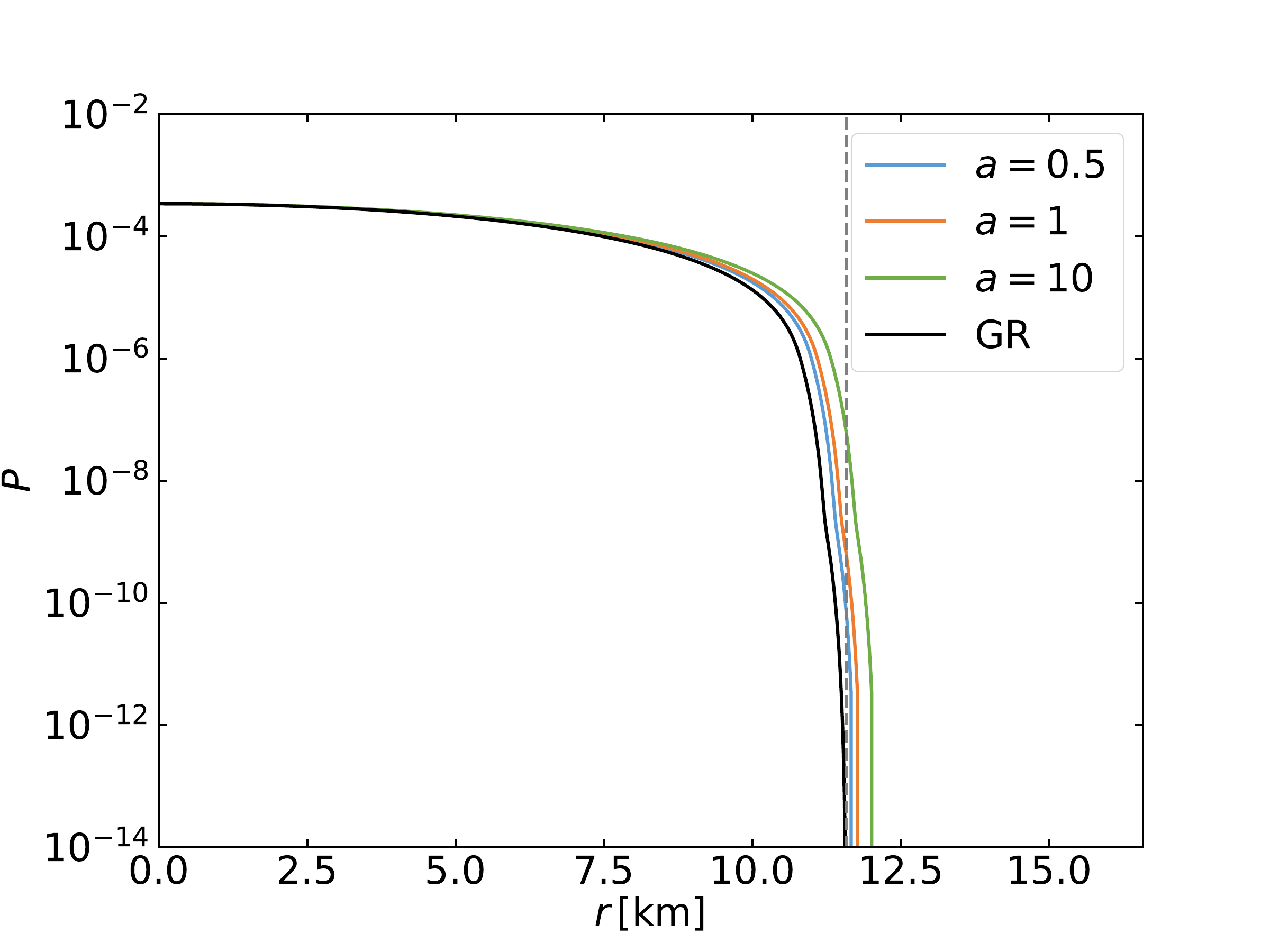}%
        \subcaption{
            Pressure profile $p(r)$
        }
    \label{fig:P_r_15}
    \end{minipage}
\end{tabular}
\caption{
     The numerical solutions for $\lambda(r)$, $\nu(r)$, $R(r)$, and $p(r)$ with $\rho_c = 1.0\times 10^{15} \rm g\,cm^{-3}$ and SLy EOS. 
     We have chosen $a = 0.5r_g, 1r_g, 10r_g$ and $b=2$. 
     The dashed vertical line corresponds to the radius of the star $r_s=11.58\,\rm km$ in GR as a reference. 
     The metric component $\lambda(r)$ and $\nu(r)$ for nonzero $a$ will recover to the Schwarzschild solution far away from $r_s$ shown in panels of (a) and (b).
}
\label{fig:geometryic_plot_JO}
\end{figure}

First, we plot the radial profiles of numerical solutions for $\lambda(r)$, $\nu(r)$, $R(r)$, and $p(r)$ in \figref{fig:geometryic_plot_JO}.
We have assumed the central pressure $p_c = 3.44\times 10^{-4}p_g$, corresponding to the central rest-mass density $\rho_c=1.0\times 10^{15}\,\rm gcm^{-3}$, with SLy EOS.
We also plot those four functions in GR ($a=0$) as the reference, 
where we impose the boundary conditions Eq.~\eqref{eq:boundary_conditions} to connect to the Schwarzschild spacetime outside the compact star.

The metric components $\lambda(r)$ and $\nu(r)$ in the NIP model show deviations from GR, while they coincide with GR predictions far away from the star as in \figref{fig:lambda_r_15} and \figref{fig:nu_r_15}.
We note that for small $a$, $\lambda(r)$ and $\nu(r)$ in the NIP model well approximate those in GR, which is consistent with GR limit ($a \rightarrow 0$) at the Lagrangian level.
For nonzero $a$, \figref{fig:R_r_15} exhibits that the curvature $R(r)$ does not vanish at the surface but decreases in power law of $r$ as distant away from the star, and increasing $a$ makes $R(r)$ decrease more slowly.


\subsection{Scalaron profiles inside and outside stars}

\begin{figure}[htbp]%
\begin{tabular}{cc}
    \begin{minipage}{0.5\linewidth}%
        \centering%
        \includegraphics[keepaspectratio, width=0.9\linewidth]{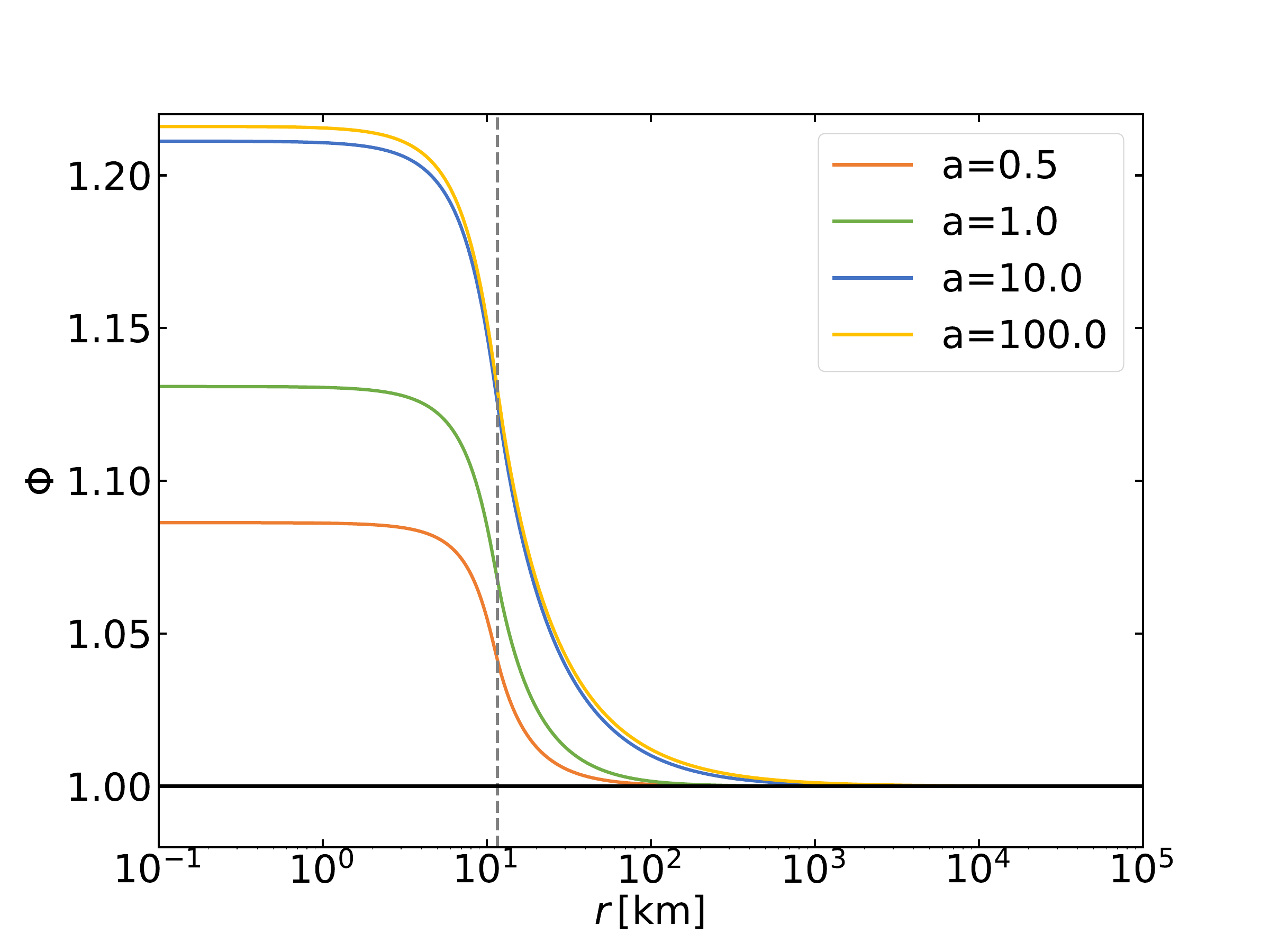}%
        \subcaption{
            Scalaron field distribution for $\Phi_{\mathrm{sol}} $
        }%
        \label{fig:scalaron_field_distribution_phi_sol}%
    \end{minipage}%
   &
    \begin{minipage}{0.5\linewidth}%
        \centering%
        \includegraphics[keepaspectratio, width=0.9\linewidth]{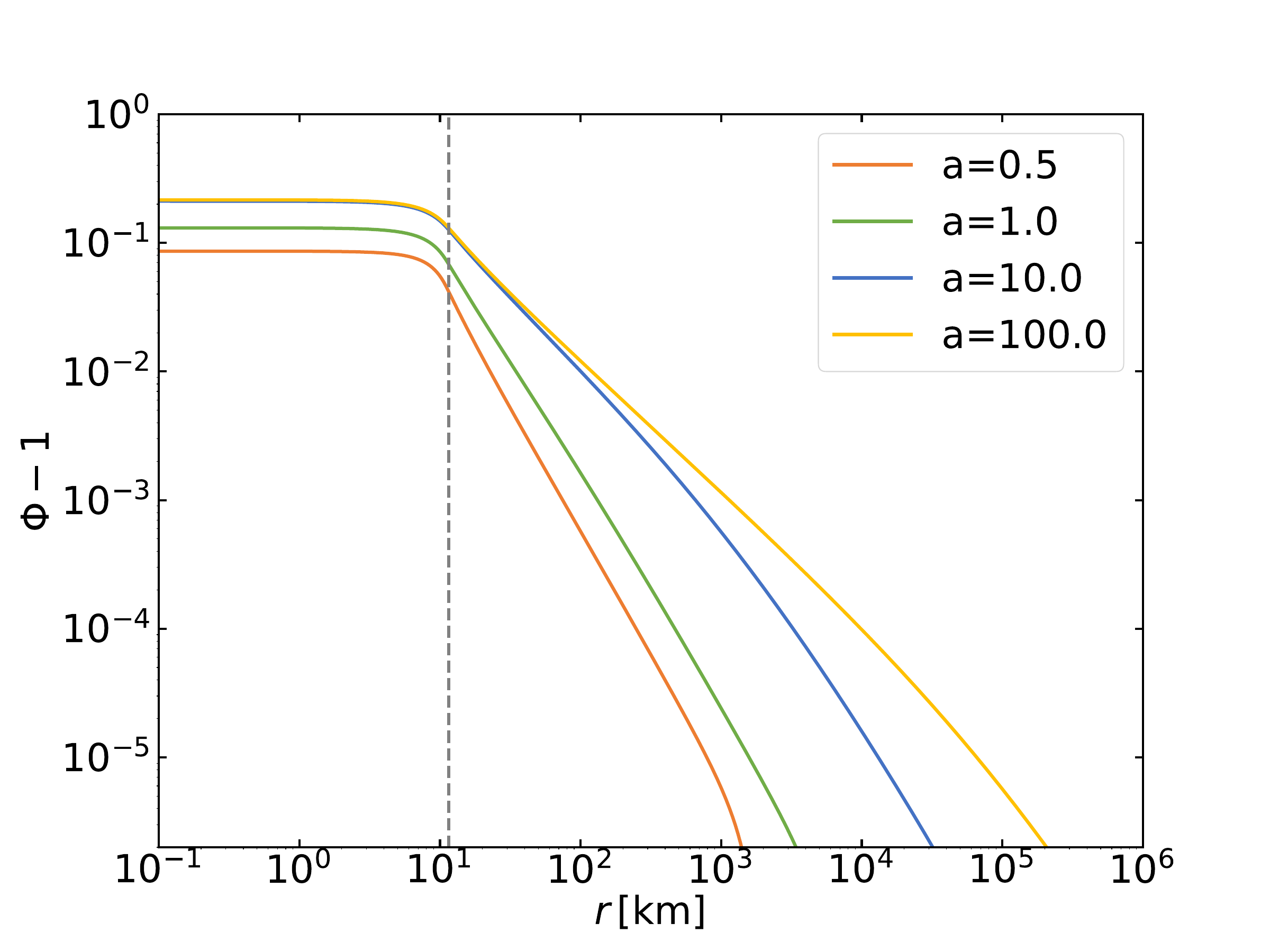}%
        \subcaption{
            Scalar hair outside the star for $\Phi_{\mathrm{sol}}$ 
        }%
        \label{fig:scalar_hair_phi_sol}%
    \end{minipage}%
    \\
    \begin{minipage}{0.5\linewidth}%
        \centering%
        \includegraphics[keepaspectratio, width=0.9\linewidth]{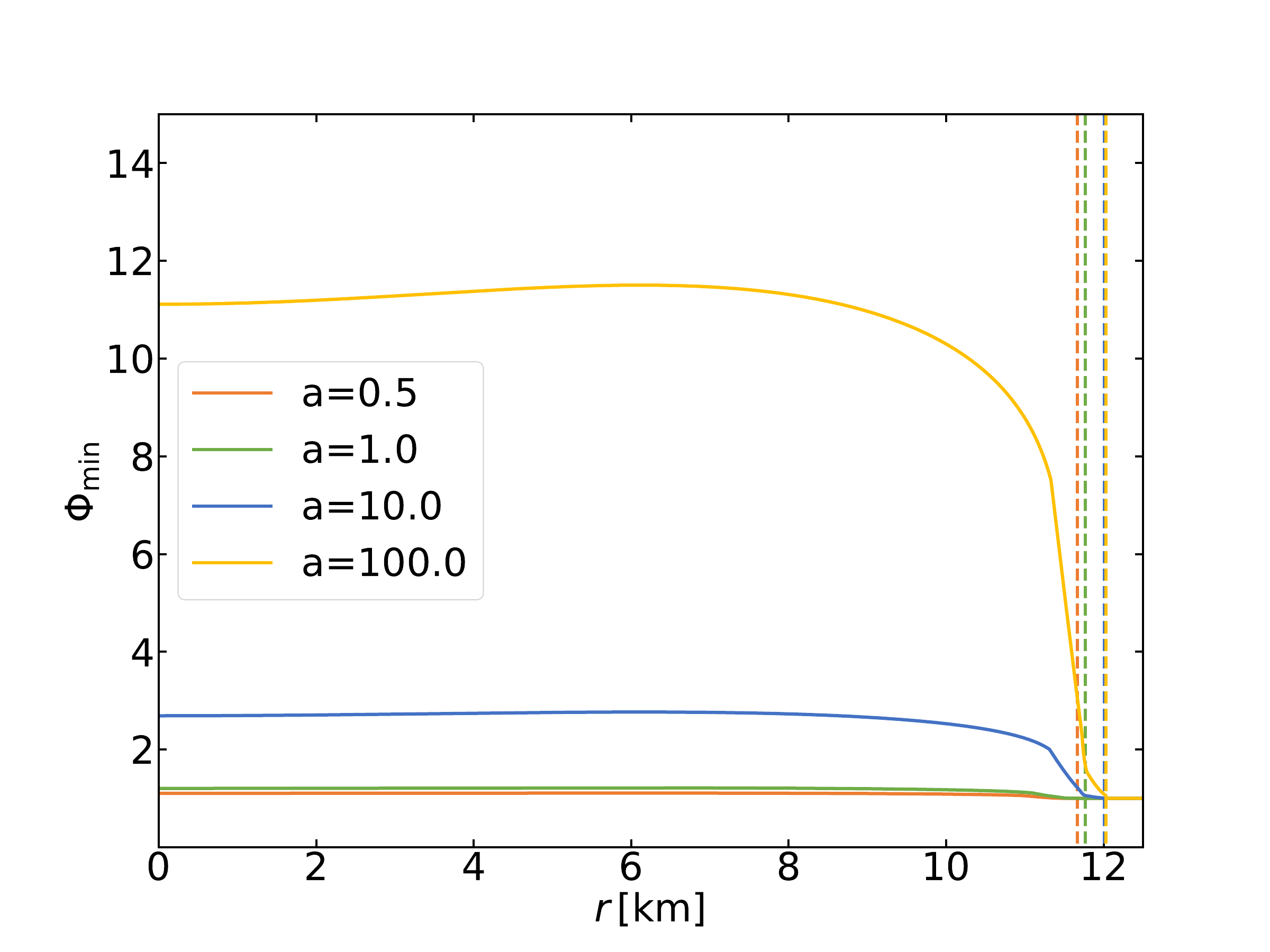}%
        \subcaption{
            Scalaron field distribution for $\Phi_{\min}$
        }%
        \label{fig:scalaron_field_distribution_phi_min}%
    \end{minipage}%
    &
    \begin{minipage}{0.5\linewidth}%
        \centering%
        \includegraphics[keepaspectratio, width=0.9\linewidth]{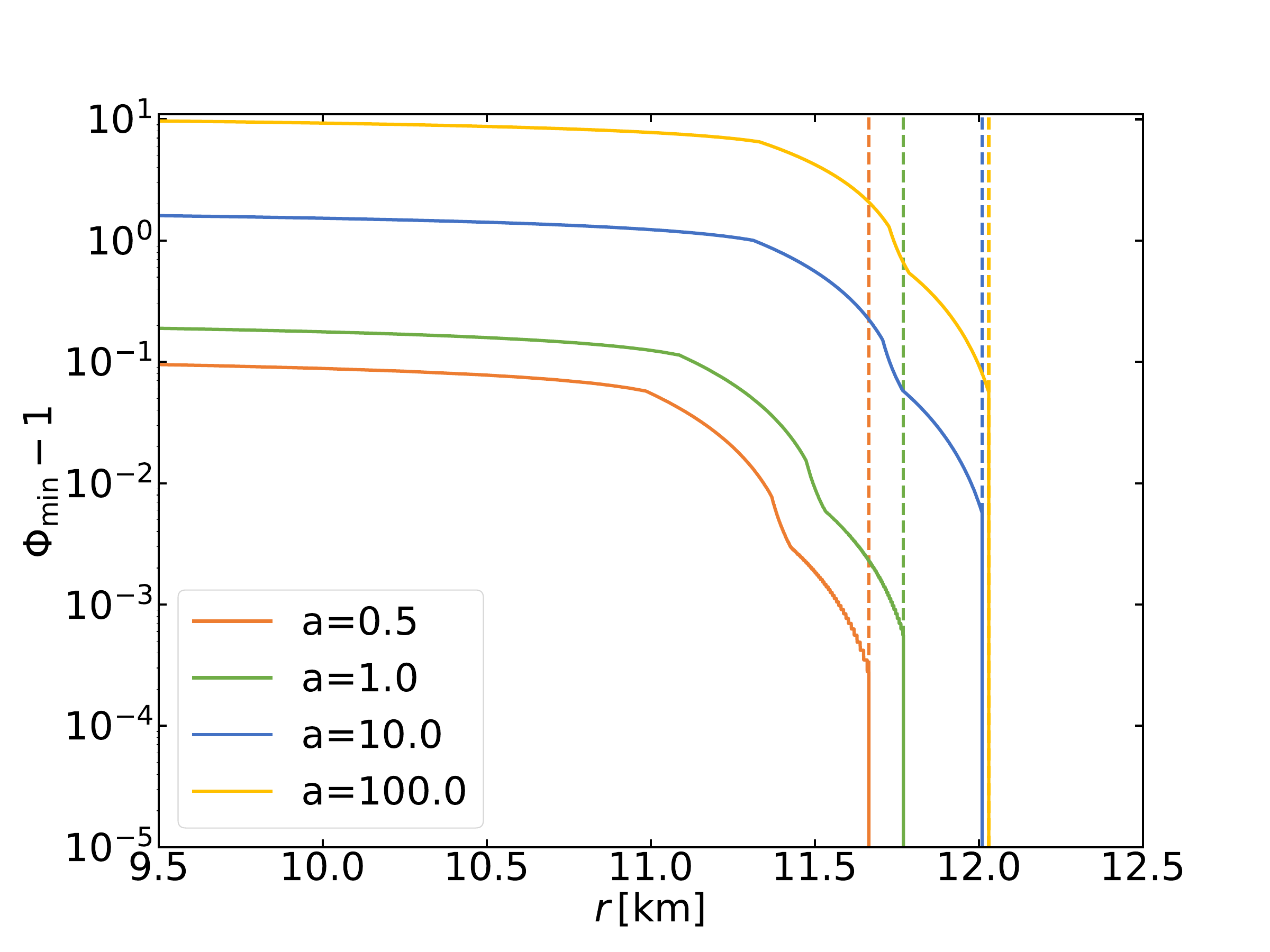}%
        \subcaption{
            Scalar hair near the star surface for $\Phi_{\min}$
        }%
        \label{fig:scalar_hair_phi_min}%
    \end{minipage}%
\end{tabular}
\caption{
    The radial profiles of scalaron field value~\eqref{eq:scalaron_def} with $\rho_c = 1.0\times 10^{15}\,\rm g\,cm^{-3}$ and SLy EOS.
    We have chosen $a= 0~(\mathrm{GR}), 0.5r_g, 1r_g, 10r_g, 100r_g$ and $b=2$. 
    The dashed vertical line and the solid horizontal line represent the radius of the star $r=r_s$ in GR and $\Phi=1$ (in GR limit), respectively. 
    The scalaron distribution with respect to $\Phi_{\mathrm{sol}}(r)$ and $\Phi_{\min}(r)$ are shown in (a) (b) and (c) (d), respectively.
}
\label{fig:scalaron_profiles}
\end{figure}

Next, we study the scalaron field $\Phi(r)$ around the compact star.
Substituting the numerically obtained solution $R(r)$ into Eq.~\eqref{eq:scalaron_def}, 
we plot the radial profile of the scalaron field $\Phi(r) = \Phi_{\mathrm{sol}}(r)$ in \figref{fig:scalaron_field_distribution_phi_sol} and the scalar hair $\Phi_{\mathrm{sol}}(r)-1$, corresponding to the deviation from GR, in~\figref{fig:scalar_hair_phi_sol}.
For the comparison, we evaluate the radial profile of the field value of the effective potential minimum $\Phi_{\min} (r)$ defined by Eq.~\eqref{eq:stationary_condition},
where we use $T(r) = - \epsilon(r) + 3p(r)$ reconstructed from the numerically obtained solution $p(r)$ and SLy EOS.
We plot $\Phi_{\min} (r)$ and $\Phi_{\min} (r)-1$ in Figs.~\ref{fig:scalaron_field_distribution_phi_min} and~\ref{fig:scalar_hair_phi_min}, respectively.

As shown in \figref{fig:scalaron_field_distribution_phi_sol},
for larger $a$, the scalaron field value $\Phi_{\mathrm{sol}}$ becomes larger inside the star. 
For any choices of parameter $a$, the scalaron field does not show drastic changes,
but it decays to $\Phi_{\mathrm{sol}}=1$ outside the star, which is the solution to the stationary condition \eqref{eq:stationary_condition} in the vacuum.
Thus, our results show the restoration of GR outside the star.
The scalar hair $\Phi_{\mathrm{sol}}-1$ also vanishes at a radius $r=r_{\Phi}$ larger than the surface $r=r_s$ as in~\figref{fig:scalar_hair_phi_sol}.
Figure~\ref{fig:scalar_hair_phi_sol} shows that the scalaron-field value decreases by a power law of $r$.
The $r_\Phi$ is prolonged as we increase $a$.
Same as $\Phi_{\mathrm{sol}}$, for larger $a$, $\Phi_{\min}$ becomes larger inside the star as shown in~\figref{fig:scalaron_field_distribution_phi_min}.
However, the difference between $\Phi_{\min}$ and $\Phi_{\mathrm{sol}}$ becomes significant for large $a$, which indicates that the scalaron field is away from the effective potential minimum.
We will discuss the above in detail later.


\subsection{Effective mass and chameleon mechanism}

\begin{figure}[htbp]%
\begin{tabular}{cc}
    \begin{minipage}{0.5\linewidth}%
        \centering%
        \includegraphics[keepaspectratio, width=0.9\linewidth]{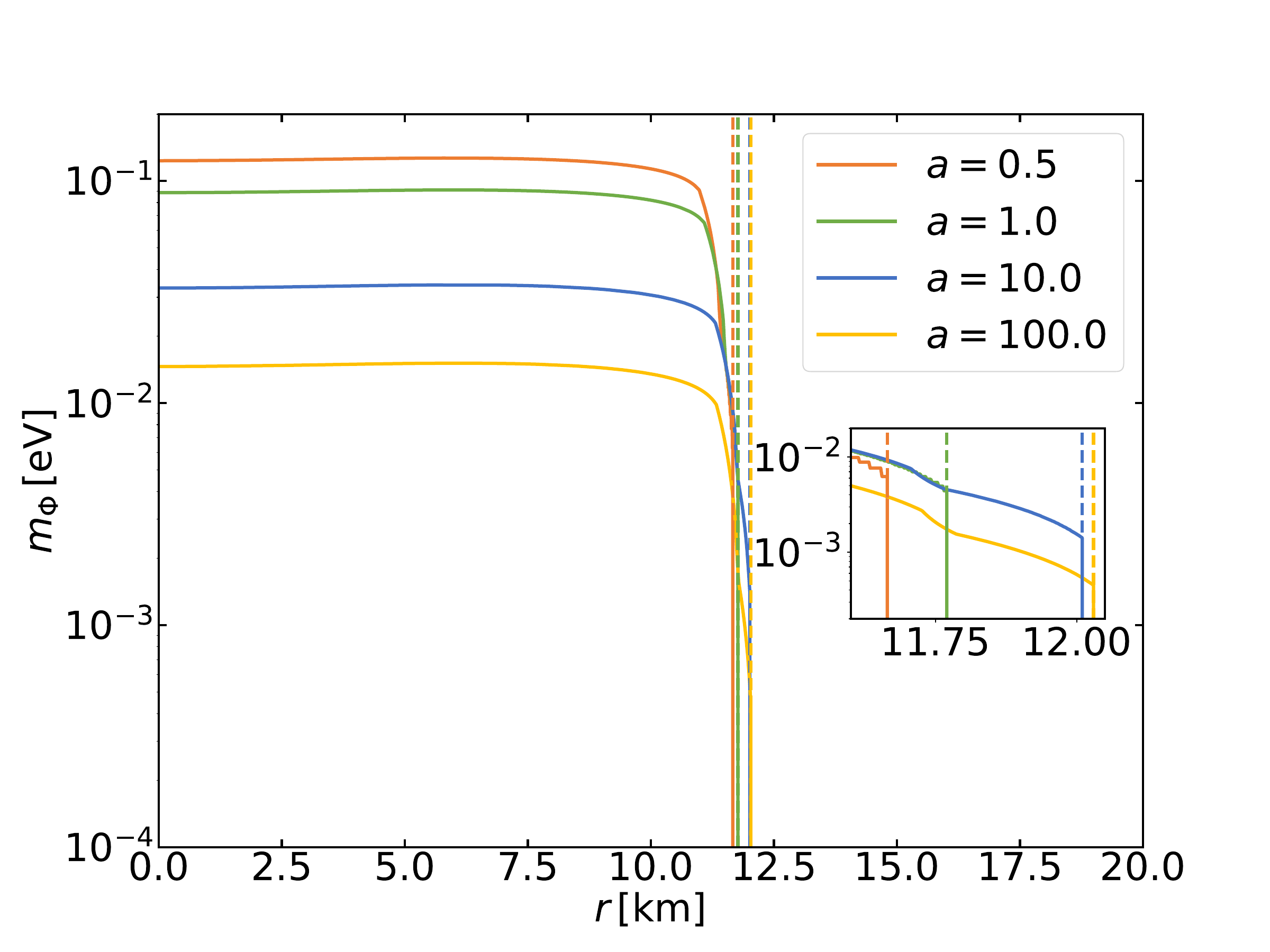}%
        \subcaption{
            The effective mass $m_{\Phi}^2$
        }%
        \label{fig:effective_mass_phi_min}%
    \end{minipage}%
    &
    \begin{minipage}{0.5\linewidth}%
        \centering%
        \includegraphics[keepaspectratio, width=0.9\linewidth]{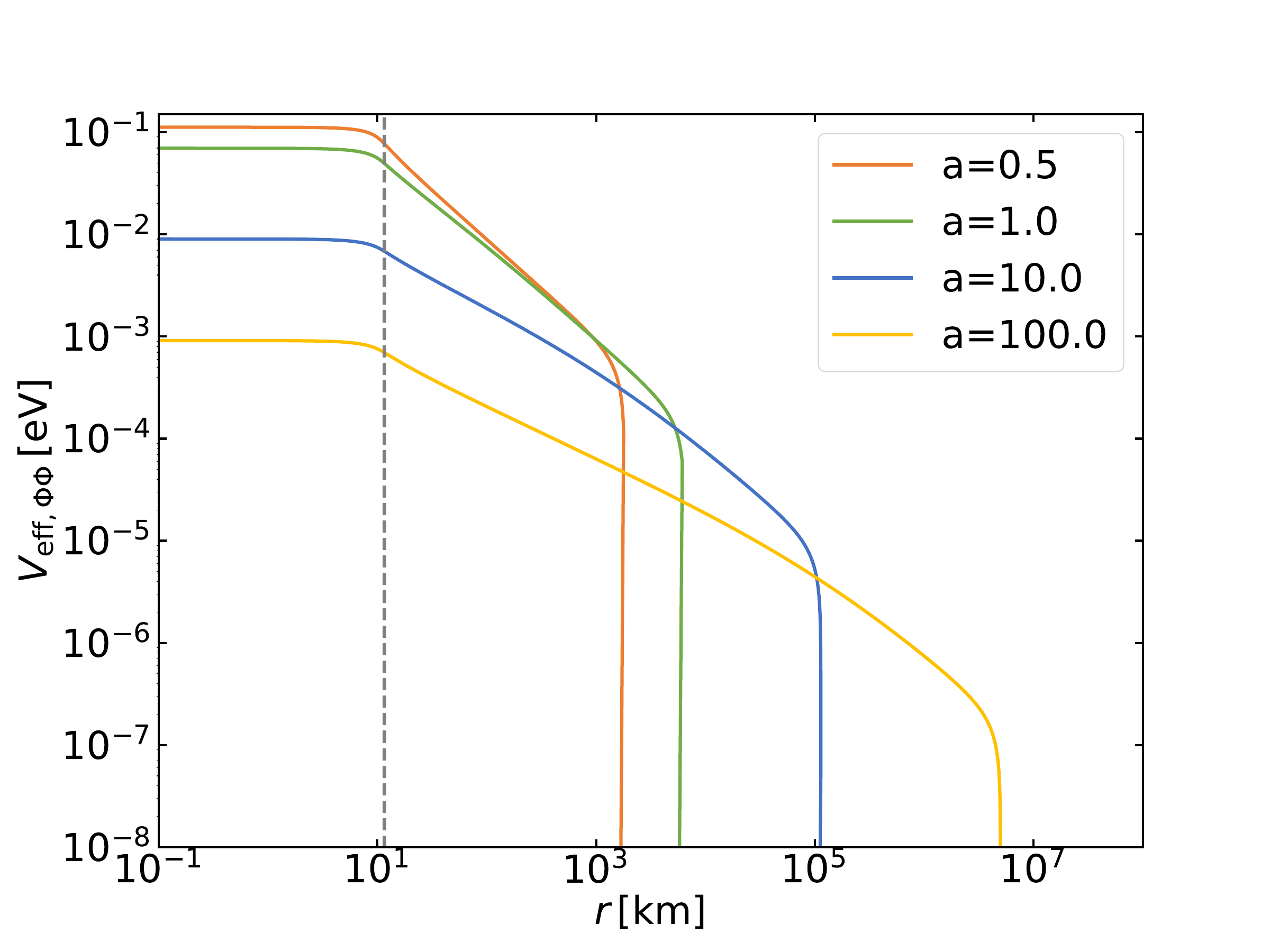}%
        \subcaption{
            $V_{\mathrm{eff}, \Phi\Phi} (\Phi)$ at $\Phi = \Phi_{\mathrm{sol}}$
        }%
        \label{fig:effective_mass_phi_sol}%
    \end{minipage}%
\end{tabular}
\caption{
    The effective mass~\eqref{eq:chameleon_Mass_NIP} of the scalaron 
    with the central rest-mass density $\rho_c=1.0\times 10^{15}\rm g\,cm^{-3}$ and SLy EOS.
    We have chosen $a = 0.5r_g, 1r_g, 10r_g, 100r_g$ and $b=2$.
    The dashed vertical line is the surface radius of the star in GR as a reference value. 
    (a) Using the stationary condition~\eqref{eq:stationary_condition} to get $\Phi_{\min}$.
    (b) Using the solution of curvature by solving the ODEs system, 
    to calculate the scalaron $\Phi_{\mathrm{sol}}$ with Eq.~\eqref{eq:scalaron_def}. 
}
\label{fig:effective_mass}
\end{figure}

In addition to the field value, we plot the scalaron effective mass~\eqref{eq:chameleon_Mass_NIP} defined by $\Phi_{\min}$ in~\figref{fig:effective_mass_phi_min}, using $T(r)$ reconstructed from the $p(r)$ and SLy EOS.
For the comparison, we also plot the value of the second derivative of the effective potential evaluated by $\Phi_{\mathrm{sol}}$ in \figref{fig:effective_mass_phi_sol}.
Note that the second quantity is not the ordinary definition of the effective mass.
Although this quantity represents the curvature of the effective potential as a function of the radial coordinate, $V_{\mathrm{eff}, \Phi\Phi}(\Phi_{\mathrm{sol}} (r))$,
the comparison with $m_{\Phi}^2(r)$ helps us understand the scalar hair and chameleon mechanism in the NIP model.

In combination with Figs.~\ref{fig:scalaron_field_distribution_phi_sol} and~\ref{fig:effective_mass_phi_min}, 
we can observe a thin-shell-like effect based on the chameleon mechanism around the compact star.
The scalaron field value takes a nearly constant value and has a large mass inside the compact star,
and $\Phi_{\mathrm{sol}}-1$ and $m_{\Phi}^2$ vanish away from the surface of the star.
Although the scalaron field is not completely screened, the field value significantly changes around the surface, and thus the gradient of the scalaron field or the fifth force is sourced by only the region around the surface of the compact star.

Our results may indicate the thin-shell effect \cite{Khoury:2003aq, Khoury:2003rn, Brax:2008hh}.
The conventional formulation of the thin-shell effect relies on the linearization of the scalaron field equation around the potential minimum $\Phi_{\min}$. 
As noticed from \figref{fig:scalaron_profiles} in the previous subsection,
the numerical solution of the scalaron field $\Phi_{\mathrm{sol}}$ significantly differs from the field value at the potential minimum $\Phi_{\mathrm{sol}}$ for the large $a$,
which cannot be handled by the perturbative approach.
Moreover, the second derivative of the potential at $\Phi = \Phi_{\mathrm{sol}}$ does not vanish outside the star, although the effective mass evaluated at the potential minimum exactly vanishes, as shown in \figref{fig:effective_mass_phi_sol}.
It is also notable that for small $a$, the scalaron field value is small with the large effective mass inside the compact star, and the scalar hair is short outside the compact star.
This is because $a \rightarrow 0$ corresponds to the GR limit, where the chameleon mechanism strongly works.

\begin{figure}[htbp]%
\begin{tabular}{cc}
    \begin{minipage}[t]{0.5\linewidth}%
        \centering%
        \includegraphics[keepaspectratio, width=0.9\linewidth]{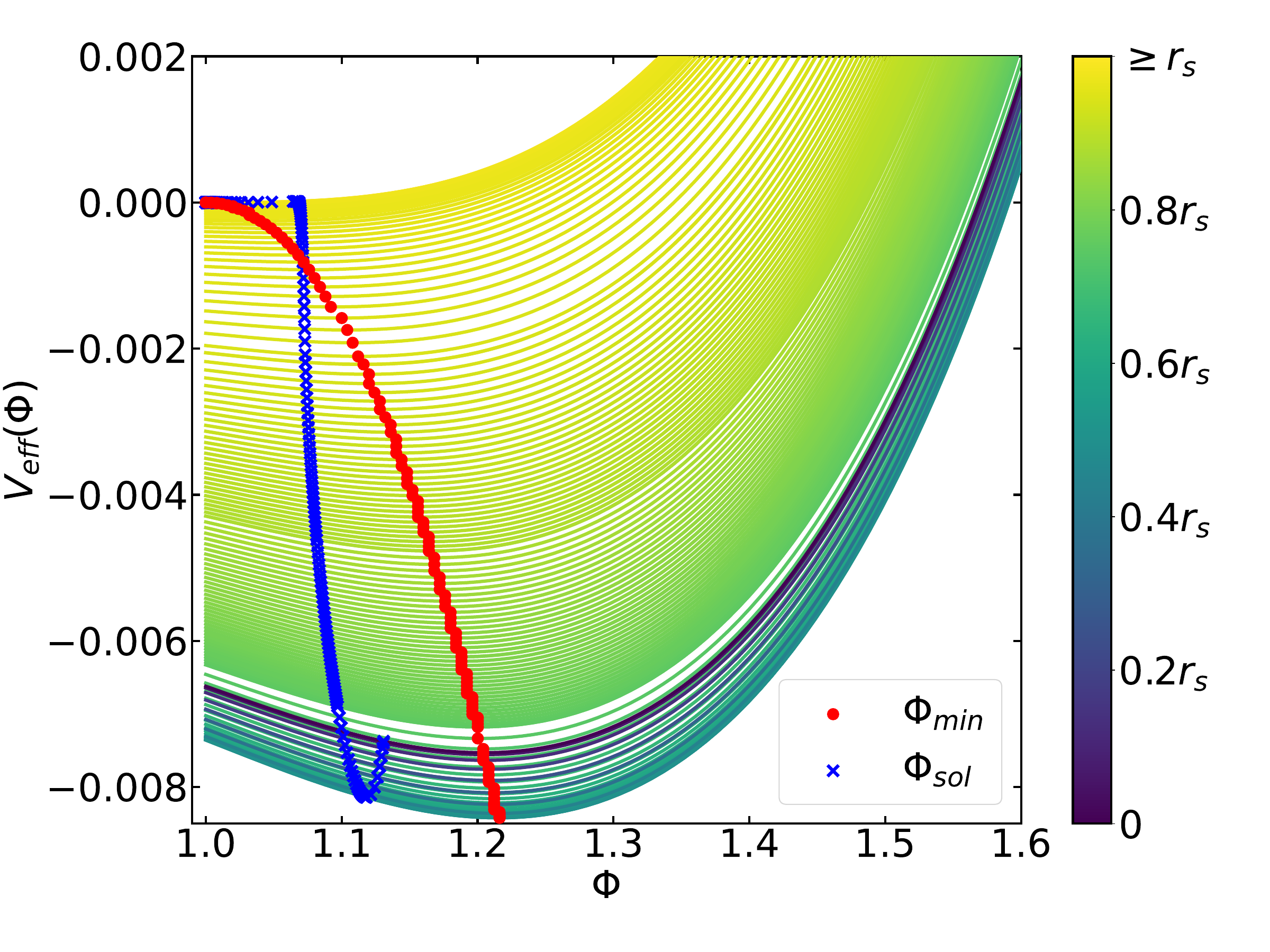}%
        \subcaption{
            The effective potential with $\Phi_{\min}$ and $\Phi_{\mathrm{sol}}$
        }%
        \label{fig:effective_pon}%
    \end{minipage}
    &
    \begin{minipage}[t]{0.5\linewidth}%
        \centering%
        \includegraphics[keepaspectratio, width=0.9\linewidth]{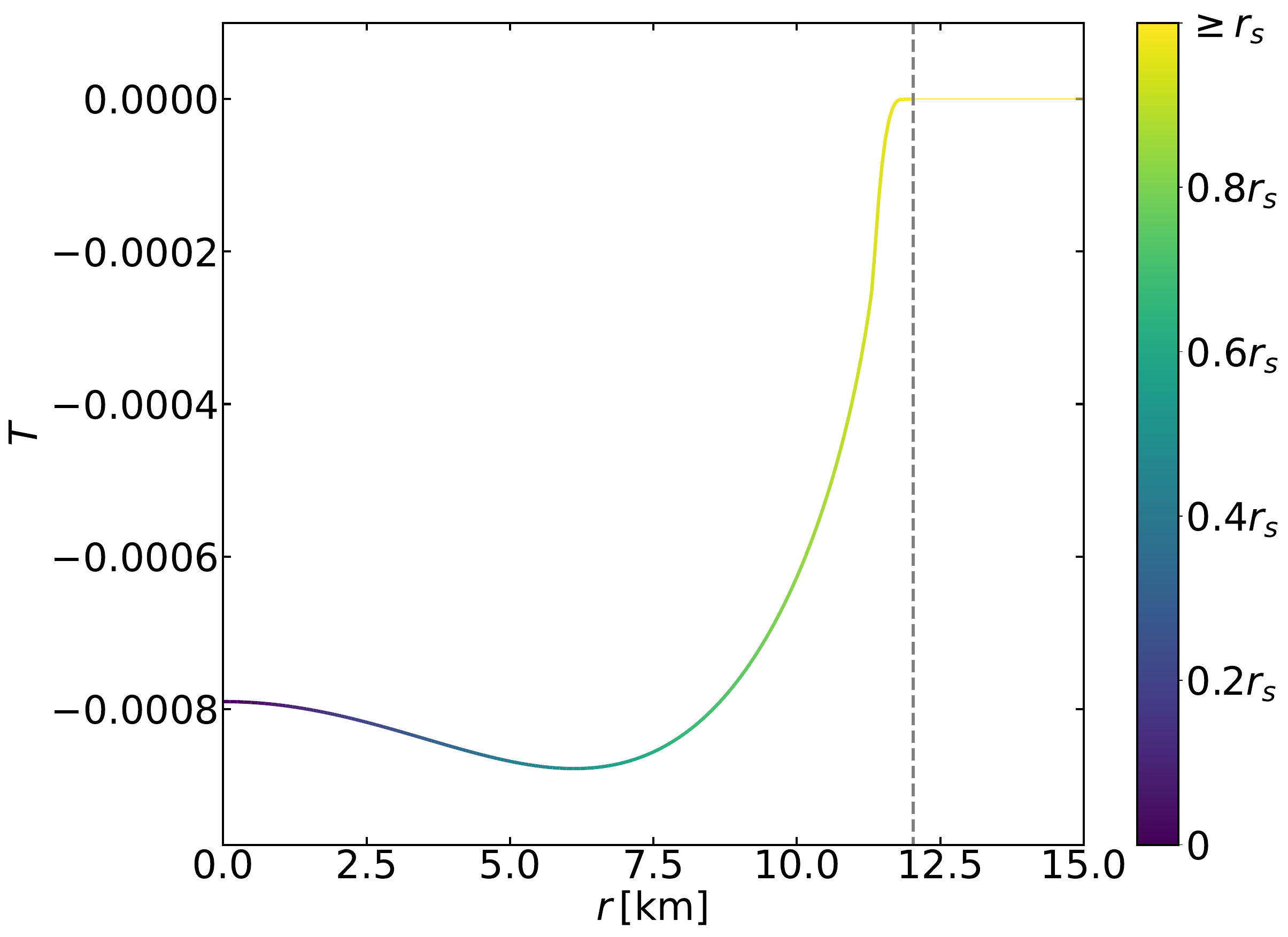}%
        \subcaption{
            The distribution of $T$
        }%
        \label{fig:T_r}%
    \end{minipage}%
\end{tabular}
\caption{
    The effective potential for scalaron controlled by T determined by the central rest-mass density $\rho_c=1.0\times 10^{15}\rm g\,cm^{-3}$ and $a=r_g$. 
    Different colored curves represent the effective potential with the different values of $T$, 
    and we treat the radial coordinate $r$ as a parameter, taking values from center $r=0$ to the surface of the star $r=r_s$. 
}
\label{fig:effective_potential_with_T}
\end{figure}

To further investigate the difference between $\Phi_{\mathrm{sol}}$ and $\Phi_{\min}$,
we plot the scalaron effective potential~\eqref{eq:definition_eff_potential} with reconstructed $T(r)$, that is, $V_{\mathrm{eff}}(\Phi(r), T(r))$ in \figref{fig:effective_pon}.
The red and blue dots on the potential correspond to locations of $\Phi_{\min}(r)$ and $\Phi_{\mathrm{sol}}(r)$, respectively. 

Figure~\ref{fig:effective_pon} shows that $\Phi_{\mathrm{sol}}$ gets across the potential minimum $\Phi_{\min}$ once inside the star and approaches $\Phi_{\mathrm{sol}}=1$ in the vacuum.  
In order to facilitate studying the change of effective potential in the radial direction, we plot the trace of the energy-momentum tensor $T$ in \figref{fig:T_r}.
One finds that $T$ approaches zero very quickly between about $10$ and $12\rm km$, which leads to the rapid change in the effective potential in this region, corresponding to the light-colored region in~\figref{fig:effective_pon}.
Note that the shape of the effective potential does not change in the vacuum ($r>r_s$).


\subsection{Mass-radius relations of compact stars}

\begin{figure}[htbp]%
\begin{tabular}{cc}
    \begin{minipage}{0.5\linewidth}%
        \centering%
        \includegraphics[keepaspectratio, width=0.9\linewidth]{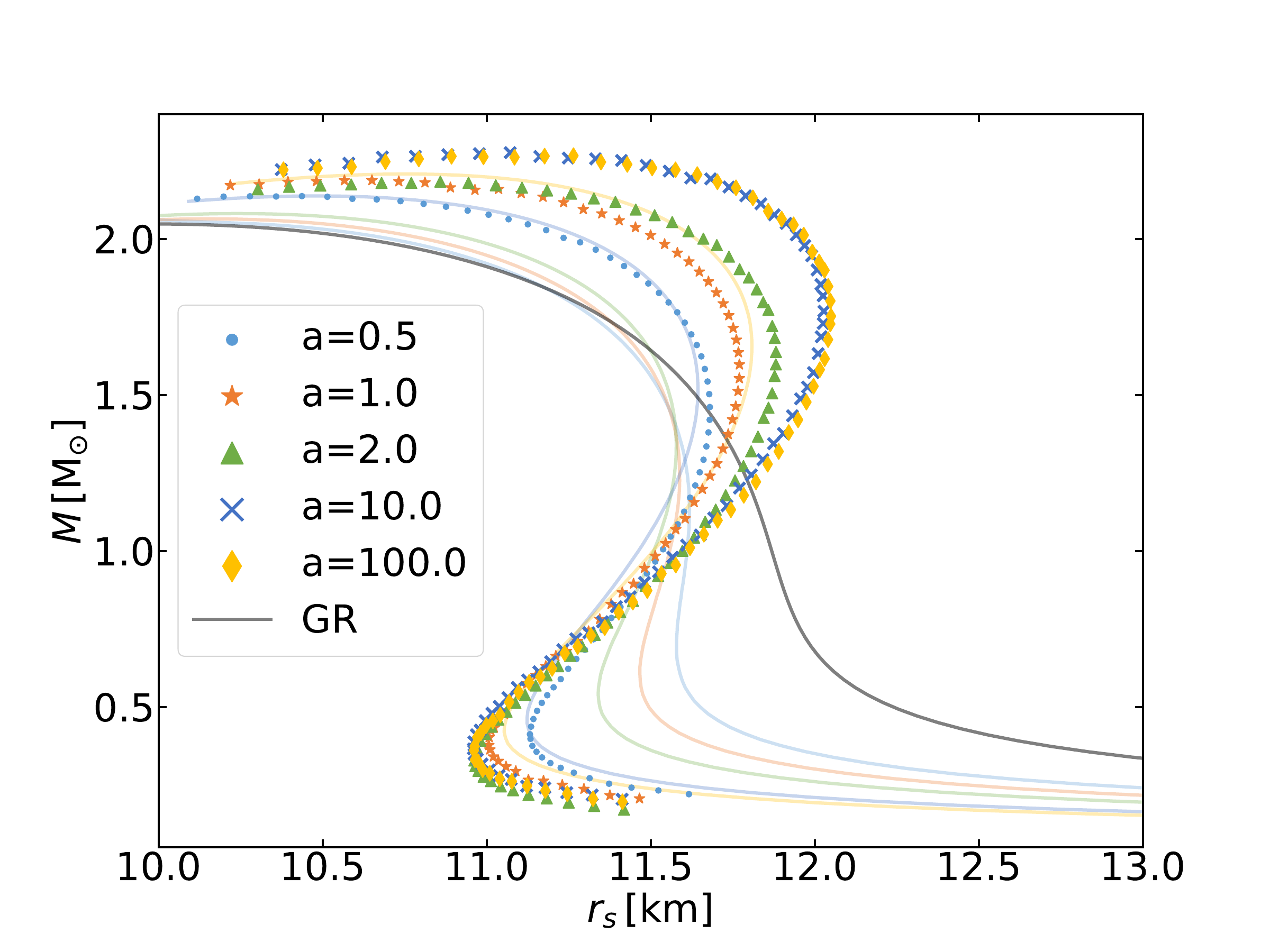}%
        \subcaption{
            M-R relation 
        }%
        \label{fig:m-r_relation}%
    \end{minipage}%
    &
    \begin{minipage}{0.5\linewidth}%
        \centering%
        \includegraphics[keepaspectratio, width=0.9\linewidth]{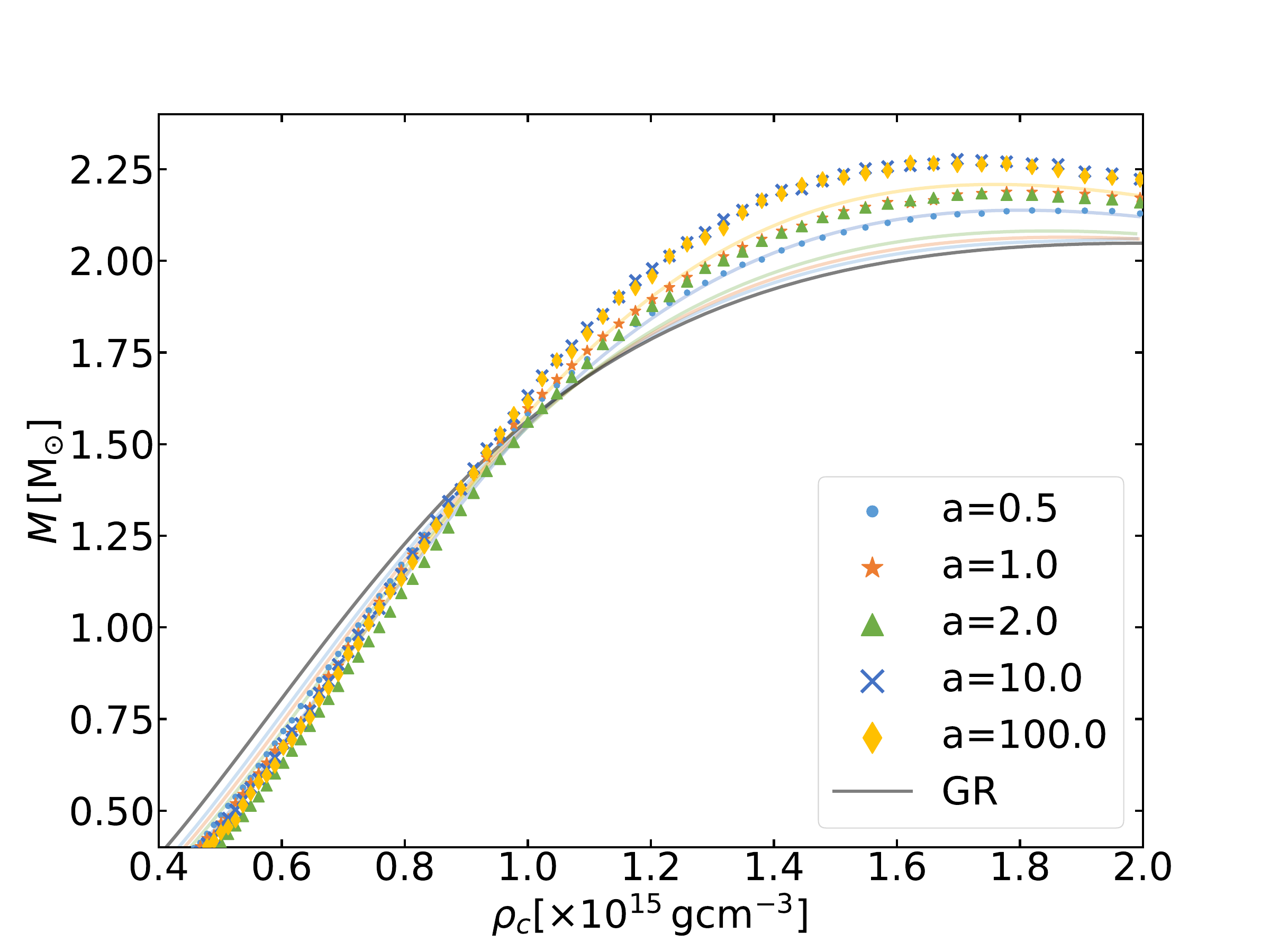}%
        \subcaption{
            Mass-$\rho_c$ relation 
        }%
        \label{fig:m_rho_relation}%
    \end{minipage}%
\end{tabular}
\caption{
    M-R relation and mass-$\rho_c$ relation constructed by the numerically obtained solutions.
    We have chosen $a = 0.5r_g, 1r_g, 10r_g, 100r_g$ and $b=2$ with SLy EOS
    The solid colored curves represent results in $R^2$ gravity~\cite{Numajiri:2023uif} as the reference.
    The parameter $a=\alpha$ is chosen for $f(R) = R + \alpha R^2$, corresponding to the limit $b\rightarrow1$.
    We impose the upper limit on $p_c$ from the condition $dp/d\epsilon<1$.
}
\label{fig:Mass-radius-density}
\end{figure}

Third, we show the M-R relation of compact stars and discuss how the scalaron contributes to it. 
We obtain the M-R relation of the compact star by calculating the mass $M$ and radius $r_s$ by changing the central rest-mass density $\rho_c$ for SLy EOS.
We plot the M-R and mass-central density relations in \figref{fig:Mass-radius-density}.
We have employed the upper limit of central pressure $p_c$ given by the EOS, 
which ensures that the speed of sound $ v^2 =\frac{dp}{d\epsilon}$ does not exceed the speed of light.

M-R curves in \figref{fig:m-r_relation} exhibit a distortion resembling a clockwise rotation around specific fixed points in response to the parameter $a$, which is similar to the case in the $R^2$ model~\cite{Numajiri:2023uif}.
As illustrated \figref{fig:m_rho_relation},
we can observe a decrease in the mass of stars with central rest-mass densities $\rho_c$ less than approximately $10^{15}~\mathrm{g\,cm^{-3}}$ and an increase in the mass of stars with $\rho_c$ greater than this value, 
This phenomenon results in the larger masses in the NIP gravity with larger $\rho_c$ compared to predictions in the GR with the same central density, 
while stars with lower $\rho_c$ have relatively reduced their masses. 
As a result, the M-R curves seem to rotate as shown in \figref{fig:m-r_relation}.

Regarding the upper limit of the central pressure,
we have other criteria to determine it.
If we demand the condition that the effective mass in the Einstein frame is always positive,
there are two possible choices of upper limits,
which leads to the truncated M-R curves to be shown in Appendix. 
In addition, the compactness $\frac{2M}{r_s}$ tends to increase for large $\alpha$.
The Buchdahl-Bondi limit gives the upper bound $\frac{2M}{r_s}<\frac{8}{9}$ for any regular and thermodynamically stable perfect fluid star in GR~\cite{Buchdahl:1959zz},
while the compactness can exceed this limit in $f(R)$ gravity theory~\cite{Goswami:2015dma}.
At least, our results with current parameter choices satisfy the original Buchdahl-Bondi limit.


\subsection{Energy conditions of scalaron field}

\begin{figure}[htbp]
\begin{tabular}{cc}
    \begin{minipage}{0.5\linewidth}%
        \centering%
        \includegraphics[keepaspectratio, width=0.9\linewidth]{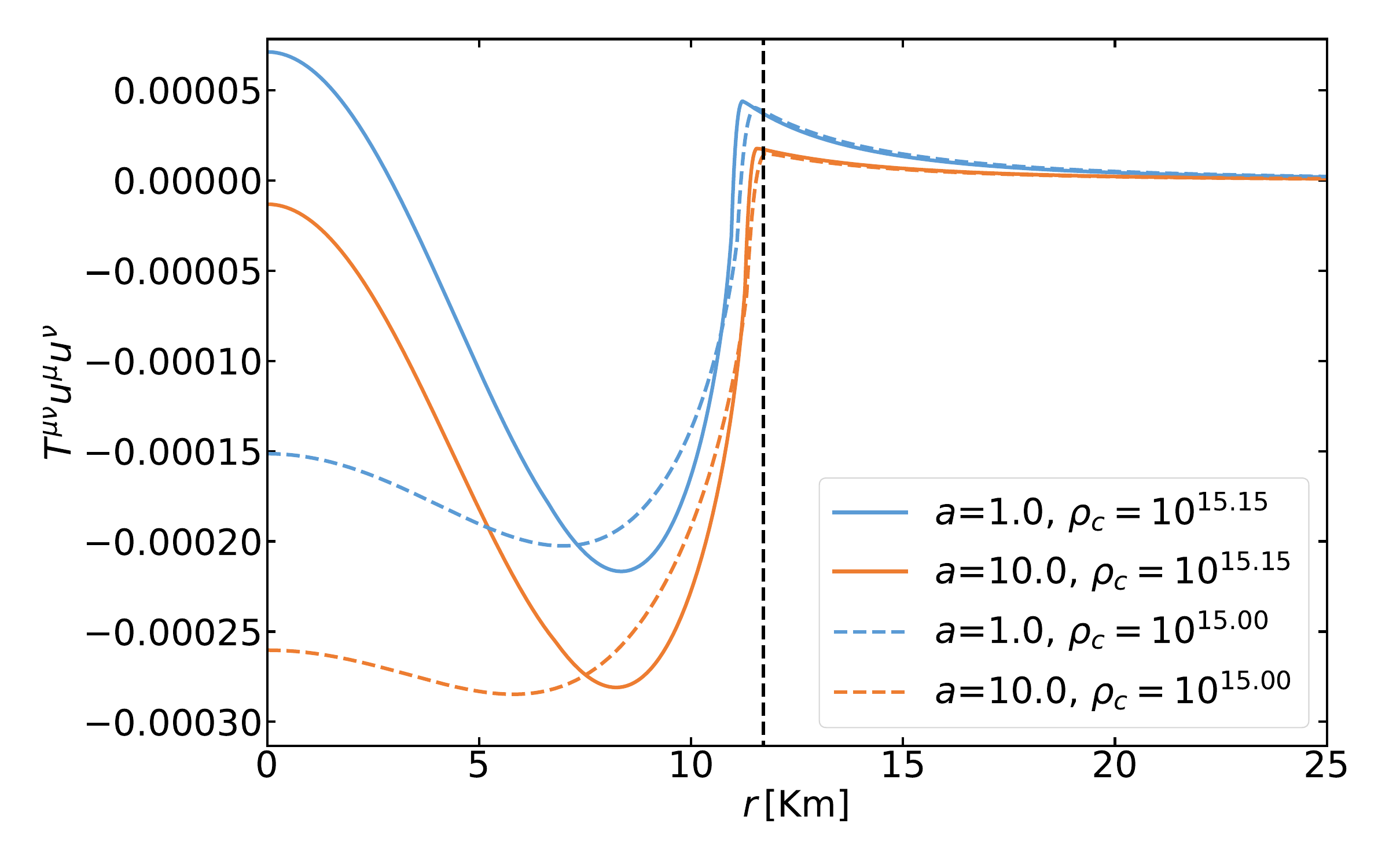}%
        \subcaption{$T^{(\Phi)}_{\mu\nu}u^{\mu}u^{\nu}$}%
        \label{fig:Tuu}%
    \end{minipage}%
    &
    \begin{minipage}{0.5\linewidth}%
        \centering%
        \includegraphics[keepaspectratio, width=0.9\linewidth]{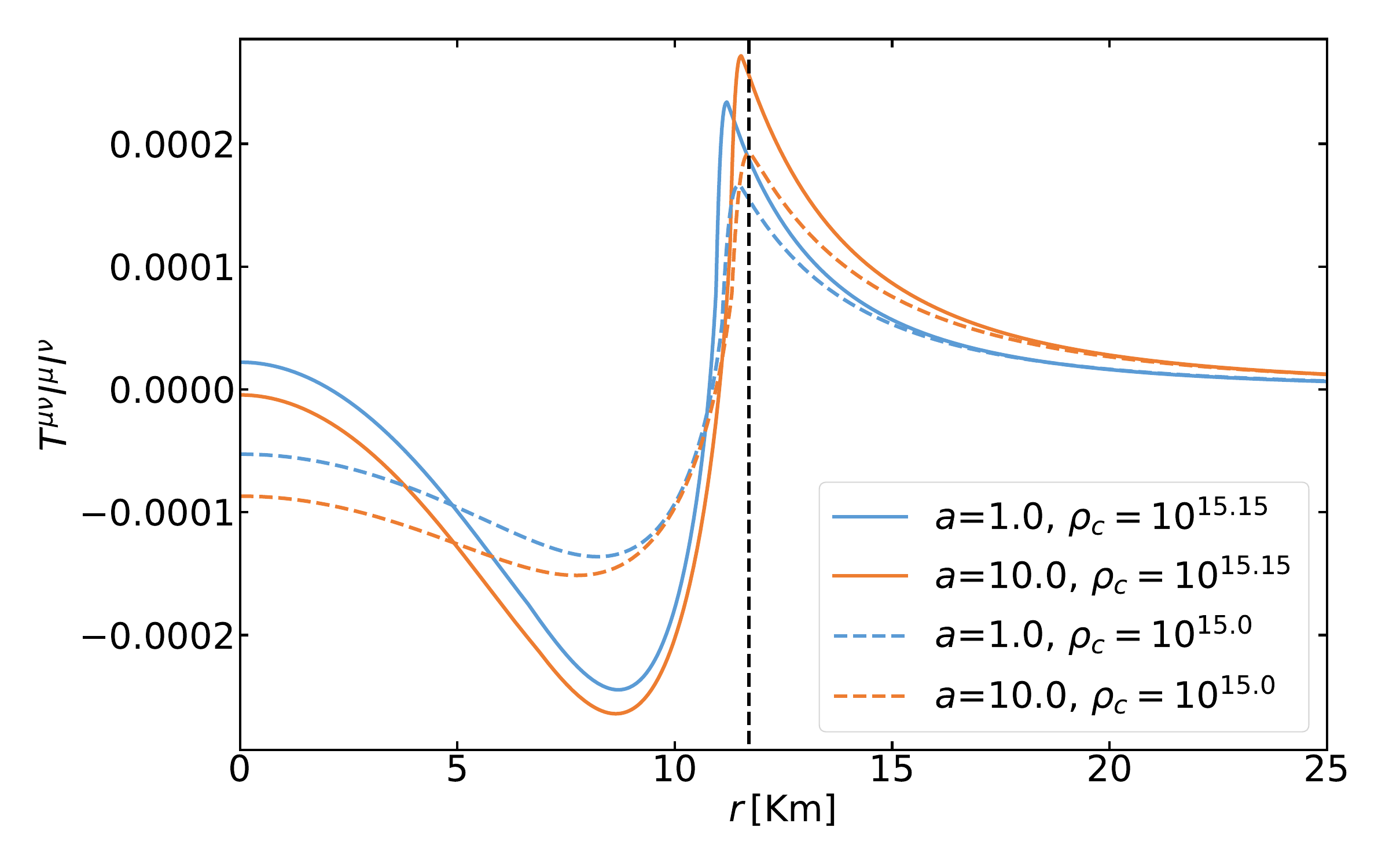}%
        \subcaption{$T^{(\Phi)}_{\mu\nu}l^{\mu}l^{\nu}$}%
        \label{fig:Tll}%
    \end{minipage}%
\end{tabular}
\caption{
    The radial profile of the energy conditions for the scalaron field.
    We have chosen $a=r_g,10 r_g$ and $\rho_c=10^{15}\mathrm{g\,cm^{-3}},10^{15.15}\mathrm{g\,cm^{-3}}$.
    The dashed vertical line is the averaged radius obtained by averaging the stellar radius for chosen parameter sets. 
}
\label{fig:energy-condition}
\end{figure}

To better understand the contribution of the scalaron field to stellar mass, 
we examine the effective energy conditions of the scalar field $\Phi$ based on the method established in Ref.~\cite{Numajiri:2023uif}. 
Evaluating the energy-momentum tensor for the scalaron field in Eq.~\eqref{eq:energy-momentum tensor for scalar},
we compute the energy conditions of the scalaron field defined by $T^{(\Phi)}_{\mu\nu}t^{\mu}t^{\nu}$ and $T^{(\Phi)}_{\mu\nu}l^{\mu}l^{\nu}$, 
where $t^{\mu}$ and $l^{\mu}$ are arbitrary timelike and null vectors, respectively. 
In this analysis, we use the four velocity of the fluid $u^{\mu}$ and the null vector $l^{\nu}=\{e^{\lambda-\nu},1,0,0\}$ for evaluation. 
We plot two energy conditions as a function of radial coordinate in \figref{fig:energy-condition}.

Figures \ref{fig:Tuu} and~\ref{fig:Tll} show that the scalar field contributes differently to the stellar mass inside and outside the star.
One can see that $T^{(\Phi)}_{\mu\nu}t^{\mu}t^{\nu}$ and $T^{(\Phi)}_{\mu\nu}l^{\mu}l^{\nu}$ are always positive outside regardless of the parameter choices.
Therefore, it can be inferred that the scalar field behaves similarly to ordinary baryonic matter outside the star, and the scalaron field will increase the stellar mass.

\begin{figure}[htbp]%
    \centering%
    \includegraphics[keepaspectratio, width=0.45\linewidth]{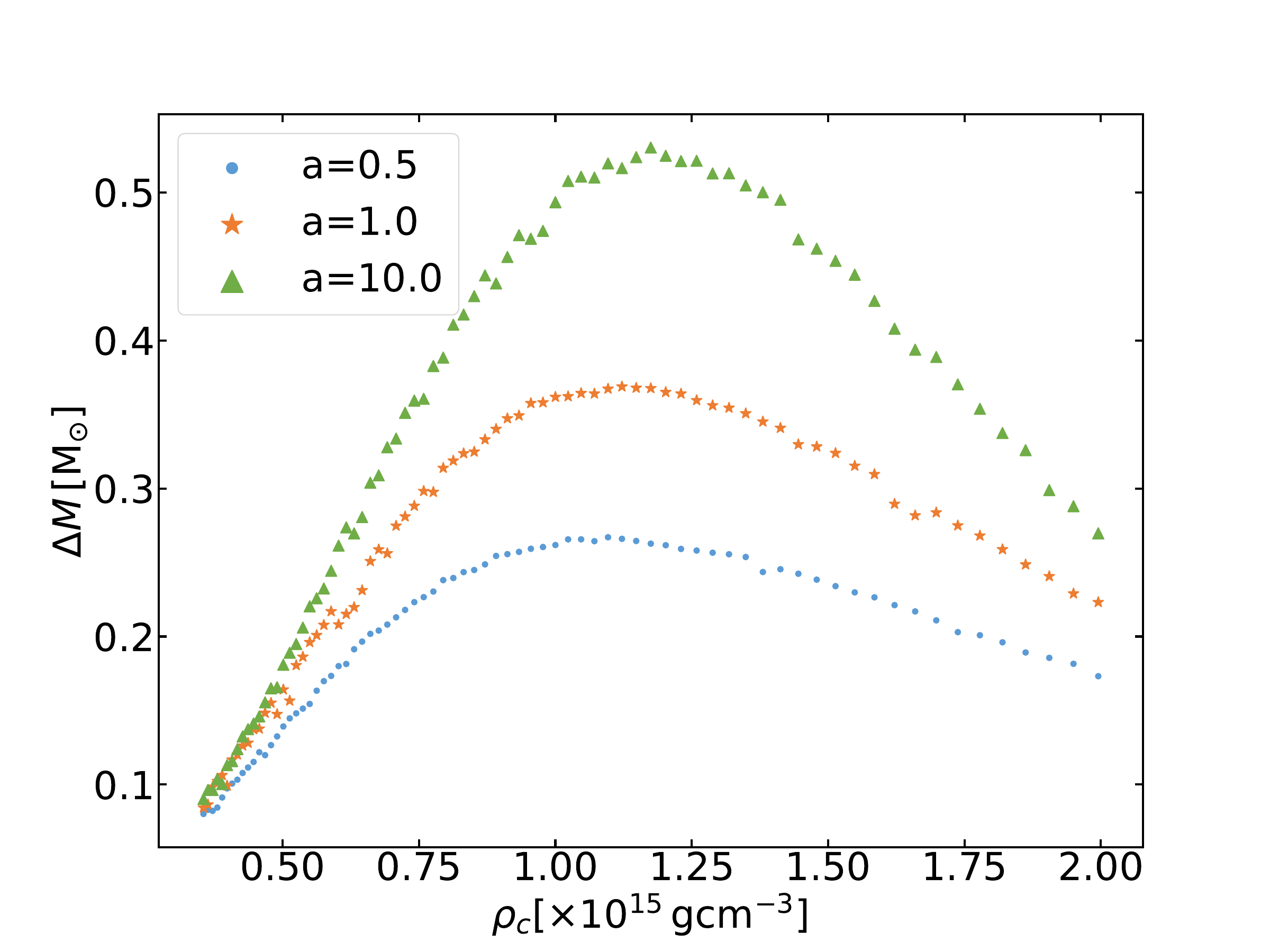}%
    \caption{
    The difference between $M$ and $m(r_s)$ with the upper limit for the possible central pressure by EOS. The figure shows the contribution of scalar hair for the observational mass $M$.
    }
\label{fig:Delta M}
\end{figure}

The scalaron contributions to the stellar mass also can be seen in $\Delta M$
\begin{align}
\label{eq:deltaM}
    \Delta M = M-m(r_s)
    \, ,
\end{align}
where $m(r_s)$ is the Schwarzschild mass which defined as Eq.~\eqref{eq:Sch_mass} on the surface $r = r_s$, and $M$ is the Schwarzschild mass on the spatial infinity $r\rightarrow\infty$. 
As shown in~\figref{fig:Delta M}, the scalaron field outside the star causes an increase in the stellar mass, and this effect becomes more pronounced as the parameter $a$ increases.

On the other hand, the contribution of the scalaron field inside the star becomes complicated due to the influence of various parameters. 
As shown in the~\figref{fig:energy-condition}, both energy conditions inside the star take nearly negative values, except for the regions near the surface and the areas near the center with high central density. 
Thus, the scalaron field $\Phi$ can violate energy conditions and behave like a quintessence field.
In other words, inside the star, the scalar field reduces the mass of the star.


\section{Conclusion}
\label{sec:conclusion}

We have studied the compact stars in the NIP model of $f(R)$ gravity, compared with known results in GR and $R^2$ model of $f(R)$ gravity.
Based on the scalar-tensor description in the Jordan frame,
we have formulated the modified TOV equations in the static and spherically symmetric spacetime and discussed the boundary conditions and asymptotic behavior of the scalaron field analytically.
We have solved the modified TOV equations numerically with the shooting method and investigated the scalaron distribution, its effective mass around the compact star, and the M-R relation of the compact star.

Regarding scalaron field distribution surrounding the compact star, the scalaron field takes a nearly constant value inside the star and starts decreasing by the power law near the surface.
This behavior can be understood in terms of the chameleon mechanism.
Although we can find similar behavior in the $R^2$ model,
the $R^2$ model does not possess the chameleon mechanism, and the scalaron field shows the exponential decay $e^{-mr}$ where $m$ represents the constant scalaron mass in the $R^2$ model.
Moreover, comparing the numerically obtained solution $\Phi_{\mathrm{sol}}$ with the field value at the effective potential minimum $\Phi_{\min}$, we have observed a significant difference between these two quantities inside and outside the compact stars.
This result indicates the nonlinear effects.

We have also considered the scalaron contributions to the stellar mass.
The evaluations of the energy condition of the scalaron field indicate that the scalaron field can behave like the quintessence field inside the compact star, while like baryonic matter outside the star.
Consequently, the scalaron field in the interior region can reduce the stellar mass significantly for lower $\rho_c$ cases, although near the surface and outside the star it increases the mass. 
These multiple contributions of the scalar field to the compact star lead to the peculiar shape of the M-R relation.
The above results are very similar to the existing results \cite{Astashenok:2021xpm, Astashenok:2021peo, Astashenok:2020qds, Astashenok:2021btj, Yazadjiev:2014cza, Astashenok:2017dpo, Astashenok:2018iav, Numajiri:2023uif} in the $R^2$ model, and it would be the common features in $f(R)$ gravity that includes the higher-curvature correction.

In addition to the scalar hair surrounding the compact star,
we have investigated the M-R relation in the NIP model, compared with the results in the GR and $R^2$ mode.
We have observed the rotation of the M-R curves, which is similar to the $R^2$ model.
However, the M-R curves in the low-mass regions largely deviate from those in the GR and $R^2$ model, which would be the characteristic nature of the NIP model.
It should be noted that we do not constrain the value of $a$ and $b$ 
because of the degeneracy problem~\cite{Numajiri:2021nsc}.
We cannot distinguish the ambiguity in EOS and that in gravitational theory in the M-R relation,
Thus, we need further observational data to separately constrain the model parameters in the $f(R)$ gravity theory.
Although we cannot clearly constrain the value of $a$ and $b$ based on the M-R relation,
the characteristic M-R relation in the low-mass region may allow us to distinguish the NIP model from GR qualitatively.


\acknowledgments

T.K. is supported by the National Key R\&D Program of China (Grant No. 2021YFA0718500)
and by Grant-in-Aid of Hubei Province Natural Science Foundation (Grant No. 2022CFB817).
K.N. is supported by JSPS KAKENHI Grant No. 23KJ1090.


\appendix

\section{SCALAR-TENSOR DESCRIPTION IN EINSTEIN FRAME}
\label{sec:EinsteinFrame}

We briefly review the NIP model in the Einstein frame.
Although we discussed the physical measurements in the Jordan frame, 
we can discuss more about the scalaron field by performing the Weyl transformation of the metric:
\begin{align}
    g_{\mu\nu} 
    & \rightarrow 
    \Tilde{g}_{\mu\nu}
    = \Phi \, g_{\mu\nu} 
    \equiv \e^{\sqrt{\frac{2}{3}}\kappa \Tilde{\Phi} } \, g_{\mu\nu}
    \, .
\end{align}
Here, $\Tilde{\Phi} $ is the scalaron field in the Einstein frame and related to $\Phi$ in the Jordan frame by
\begin{align}
\label{eq:scalarontrans}
    \Phi = \exp[\sqrt{\frac{2}{3}}\kappa \Tilde{\Phi}]
    \, .
\end{align}
According to the Weyl transformation,
the action Eq.~\eqref{eq:action_scalar_field_Jordan} is rewritten as
\begin{align}
\label{eq:action_Einstein}
    S = 
    \int d^{4} x \sqrt{-\Tilde{g}}
    \qty[\frac{\Tilde{R}}{2 \kappa^{2}} 
    - \frac{1}{2} \partial^{\alpha} \Tilde{\Phi} \partial_{\alpha} \Tilde{\Phi}
    -U(\Tilde{\Phi})]
    \, .
\end{align}
The scalaron potential in the Einstein frame $U$ is defined as
\begin{align}
    U(\Tilde{\Phi} ) 
    = \frac{1}{2\kappa^2}\e^{-2\sqrt{\frac{2}{3}}\kappa \Tilde{\Phi} } Y(\Tilde{\Phi} )
\end{align}
This action comprises the Einstein-Hilbert action and minimally coupled canonical scalar field with potential $U(\Tilde{\Phi})$. 
The whole DOF of the gravitational field is apparently $ 2+1 $. 
This frame is often called the Einstein frame in contrast with the Jordan frame with the action Eq.~\eqref{eq:action_Jordan}. 

By variation of the gravitational action~\eqref{eq:action_Einstein} and the matter action,
we can derive the field equations for $\Tilde{g}_{\mu\nu}$ and $\Tilde{\Phi}$
\begin{align}
    &\Tilde{R}_{\mu\nu} - \frac{1}{2} \Tilde{R} \Tilde{g}_{\mu\nu}
    = \kappa^2 \qty(\Tilde{T}_{\mu\nu} + \Tilde{T}^{\Tilde{\Phi}}_{\mu\nu}),
     \\
    &\Tilde{\square} \Tilde{\Phi}
    =
    U_{\Tilde{\Phi}}(\Tilde{\Phi})
    + \frac{\kappa}{\sqrt{6}} \Tilde{T} 
    \, ,
\end{align}
where $\Tilde{T}^{\Tilde{\Phi}}_{\mu\nu}$ and $\Tilde{T}_{\mu \nu}$ are defined as
\begin{align}
    \Tilde{T}^{\Tilde{\Phi}}_{\mu\nu}
    =
    \partial_{\mu} \Tilde{\Phi} \partial_{\nu} \Tilde{\Phi}
    - \Tilde{g}_{\mu\nu}
     \qty[
        \frac{1}{2} \partial^{\alpha} \Tilde{\Phi} \partial_{\alpha} \Tilde{\Phi}
        -U(\Tilde{\Phi}) 
    ]
    \\
    \Tilde{T}_{\mu \nu} \equiv \frac{2}{\sqrt{-\Tilde{g}}} \frac{\delta S_M}{\delta \Tilde{g}^{\mu \nu}}
    =\e^{-\sqrt{\frac{2}{3}}\kappa\Tilde{\phi}} \: T_{\mu\nu} 
    \,.
\end{align}
The tilde denotes the quantities in the Einstein frame.

One can also discuss the chameleon mechanism in the Einstein frame. 
The effective potential $U_{\mathrm{\mathrm{eff}}} (\Tilde{\Phi}, \Tilde{T})$ in the Einstein frame is defined as
\begin{align}
\begin{split}
    \Tilde{\square} \Tilde{\Phi}
    &= 
    U_{\Tilde{\Phi}} (\Tilde{\Phi}) +\frac{\kappa}{\sqrt{6}} \Tilde{T} 
    \\
    &= \frac{1}{\sqrt{6}\kappa}
    \qty[\frac{2F(R(\Tilde{\Phi}))-R(\Tilde{\Phi})F_R(R(\Tilde{\Phi}))+ \kappa^2 T}{F_R^2 (R(\Tilde{\Phi}))} ] 
    \\
    & \equiv 
    \pdv{U_{\mathrm{\mathrm{eff}}}}{\Tilde{\Phi}} \qty(\Tilde{\Phi}, \Tilde{T})
    \, ,
\end{split}
\end{align}
Note that $T$ without tilde denotes the Jordan frame quantity.
The chameleon mass of the scalar field in the Einstein frame is
\begin{align}
\label{eq:definition_eff_mass_Einstein}
\begin{split}
    m_{\Tilde{\Phi}}^2 
    &\equiv 
    \eval{\pdv[2]{U_{\mathrm{\mathrm{eff}}}}{\Tilde{\Phi}}}_{\Tilde{\Phi}=\Tilde{\Phi}_{\min}} 
    \\
    &=
    \frac{1}{3 F_R\left(R_{\min }\right)}
    \left(
        \frac{F_R\left(R_{\min }\right)}{F_{R R}\left(R_{\min }\right)} - R_{\min}
    \right)
    \\
    &= 
    \frac{1}{F_R\left(R_{\min }\right)} m_{\Phi}^2 
    \, ,
\end{split}
\end{align}
where $R_{\min} = R(\Tilde{\Phi}_{\min})$ defined at the potential minimum. 
The dependence on the energy-momentum tensor comes into the chameleon mass via $R_{\min}$ implicitly. 
We also note that Eq.~\eqref{eq:definition_eff_mass_Einstein} is different from Eq.~\eqref{eq:definition_eff_mass1} by the factor $F_{R}$.

\begin{figure}[tbp]%
\begin{tabular}{cc}
    \begin{minipage}[t]{0.5\linewidth}%
        \centering%
        \includegraphics[keepaspectratio, width=0.9\linewidth]{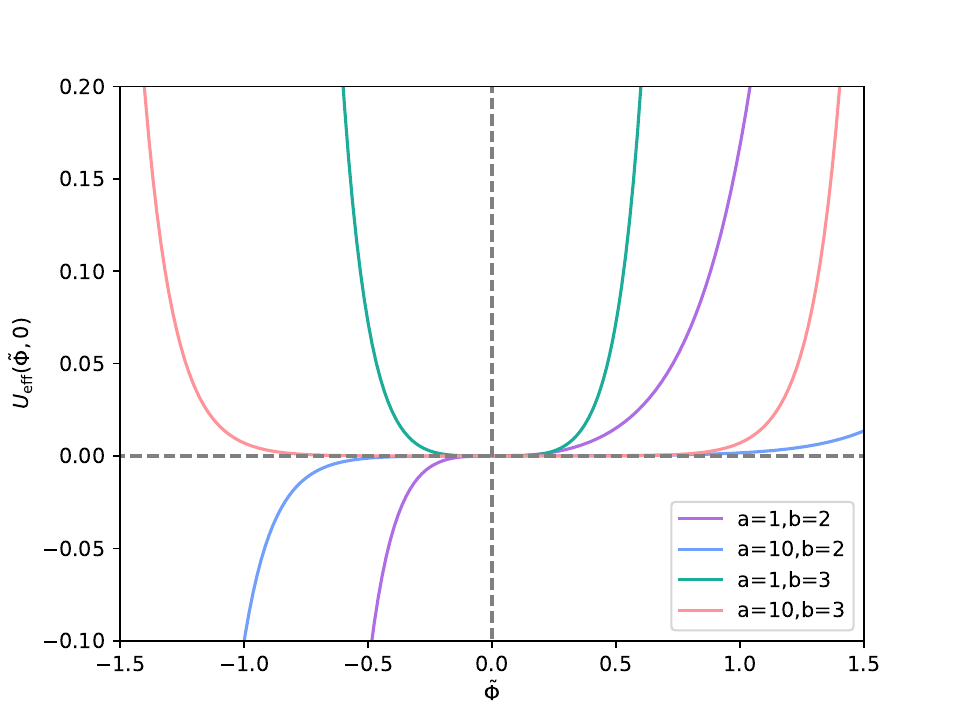}%
        \subcaption{
            Varying $a$ and $b$ with $T=0$.
        }%
        \label{fig:chame_pot_EI_0}%
    \end{minipage}%
    \begin{minipage}[t]{0.5\linewidth}%
        \centering%
        \includegraphics[keepaspectratio, width=0.9\linewidth]{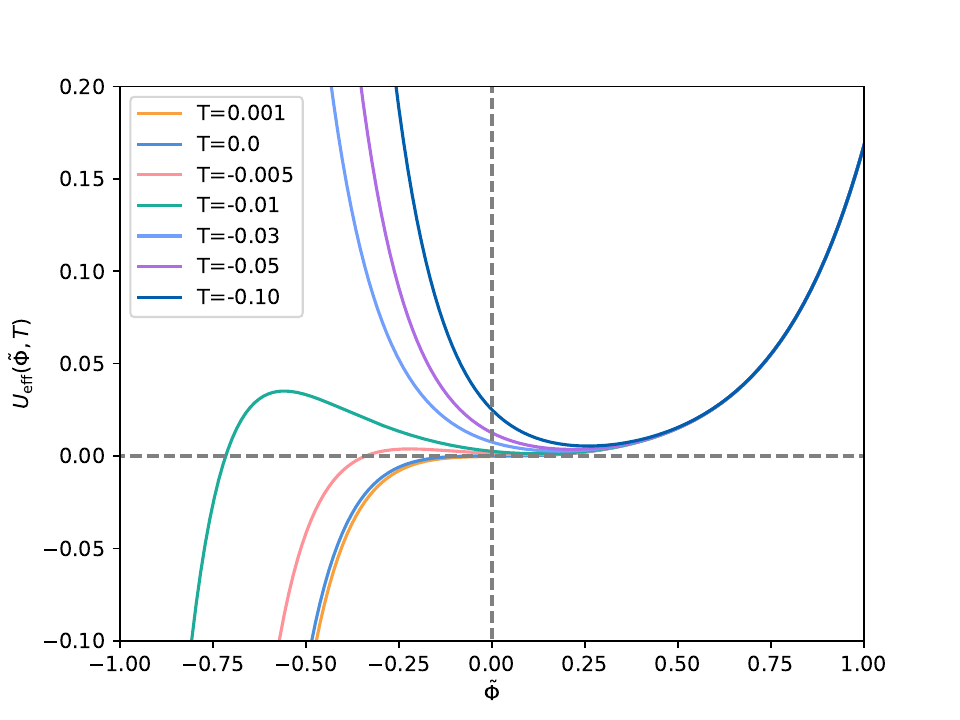}
        \subcaption{
            Varying $T$ with $a=1r_g, b=2$.
        }%
        \label{fig:chame_pot_EI_T}%
    \end{minipage}%
\end{tabular}
\caption{
    The scalaron effective potential \eqref{eq:chame_pot_Einstein} of the NIP gravity \eqref{eq:NIP_gravity} in the Einstein frame, normalized by $r_g$.
    $T$ is defined in the Jordan frame.
    (a) The bottom of the potential becomes shallower as $a$ and $b$ also increase in this frame. 
    (b) The potential minimum moves horizontally and becomes shallower as $T$ increases.
}
\label{fig:chame_pot_EI}
\end{figure}%

\begin{figure}[tbp]%
\begin{tabular}{cc}
    \begin{minipage}{0.5\linewidth}%
        \centering%
        \includegraphics[keepaspectratio, width=0.9\linewidth]{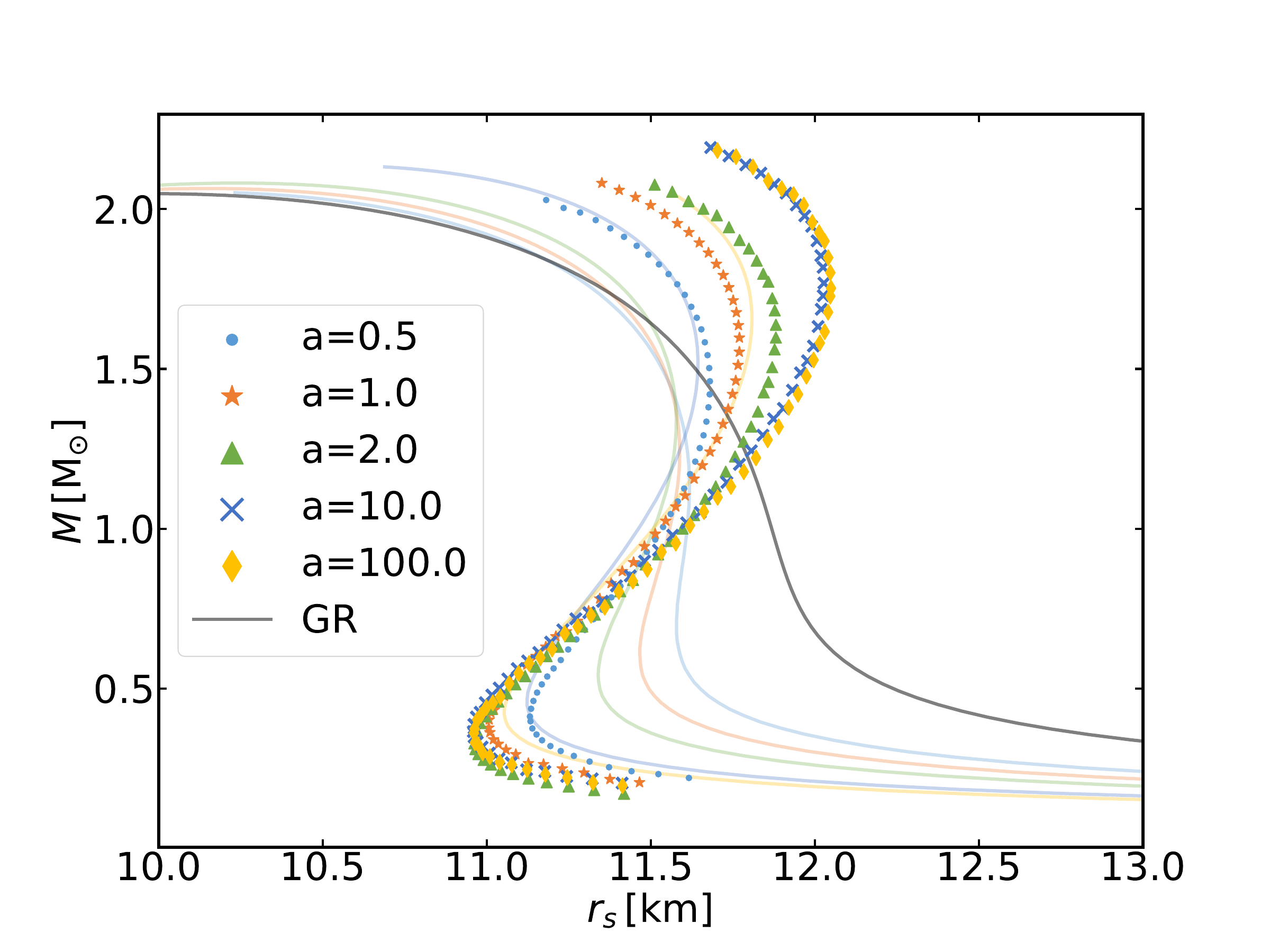}%
        \subcaption{
            Mass-$r_s$ relation
            with $p_c$ constrained by $m_{\Tilde{\Phi}}^2>0$
        }%
        \label{fig:m-r_relation_for_phi_min}%
    \end{minipage}%
    &
    \begin{minipage}{0.5\linewidth}%
        \centering%
        \includegraphics[keepaspectratio, width=0.9\linewidth]{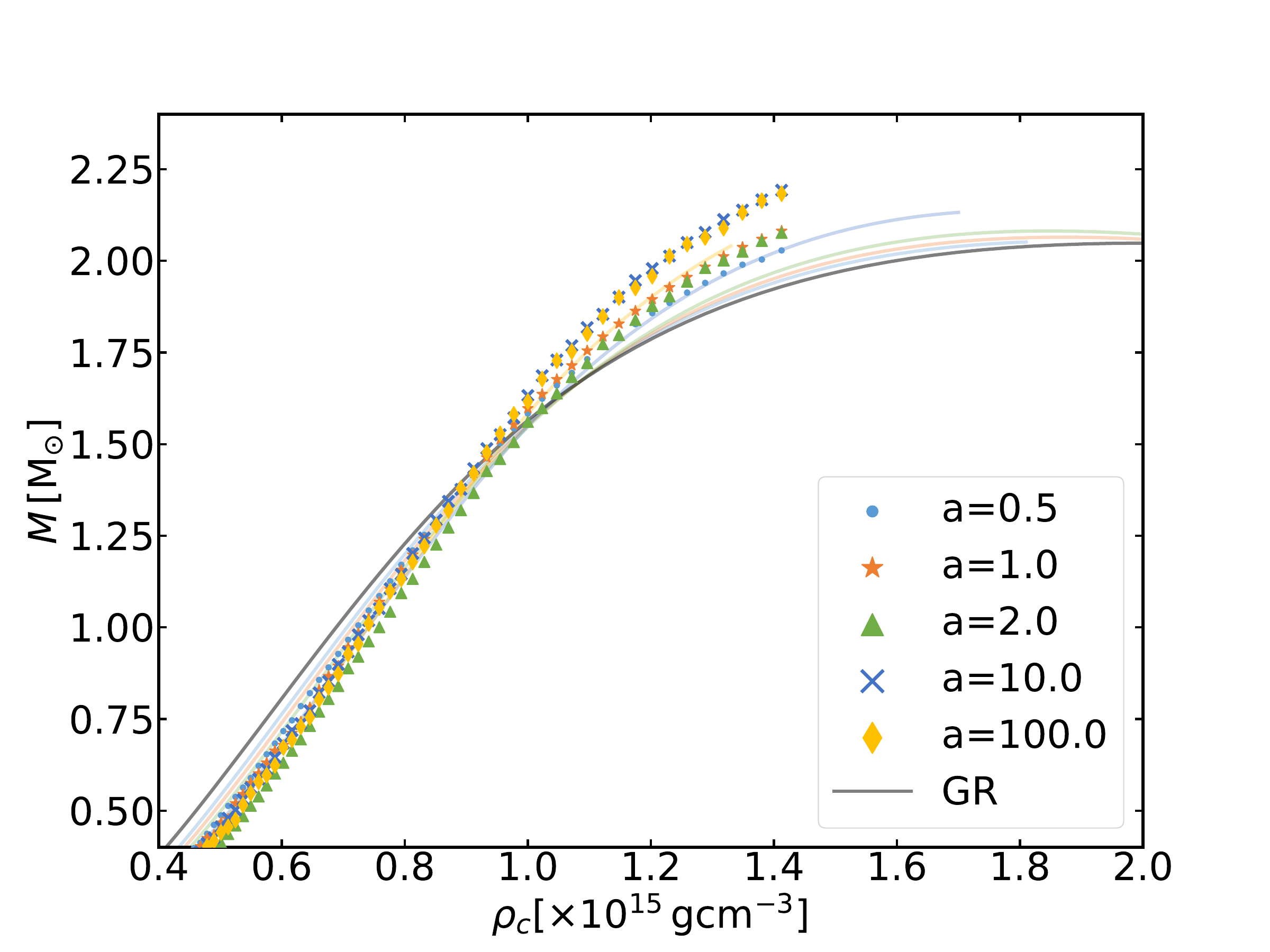}%
        \subcaption{
            Mass-$\rho_c$ relation
            with $p_c$ constrained by $m_{\Tilde{\Phi}}^2>0$
        }%
        \label{fig:m_rho_relation_for_phi_min}%
    \end{minipage}%
    \\
    \begin{minipage}{0.5\linewidth}%
        \centering%
        \includegraphics[keepaspectratio, width=0.9\linewidth]{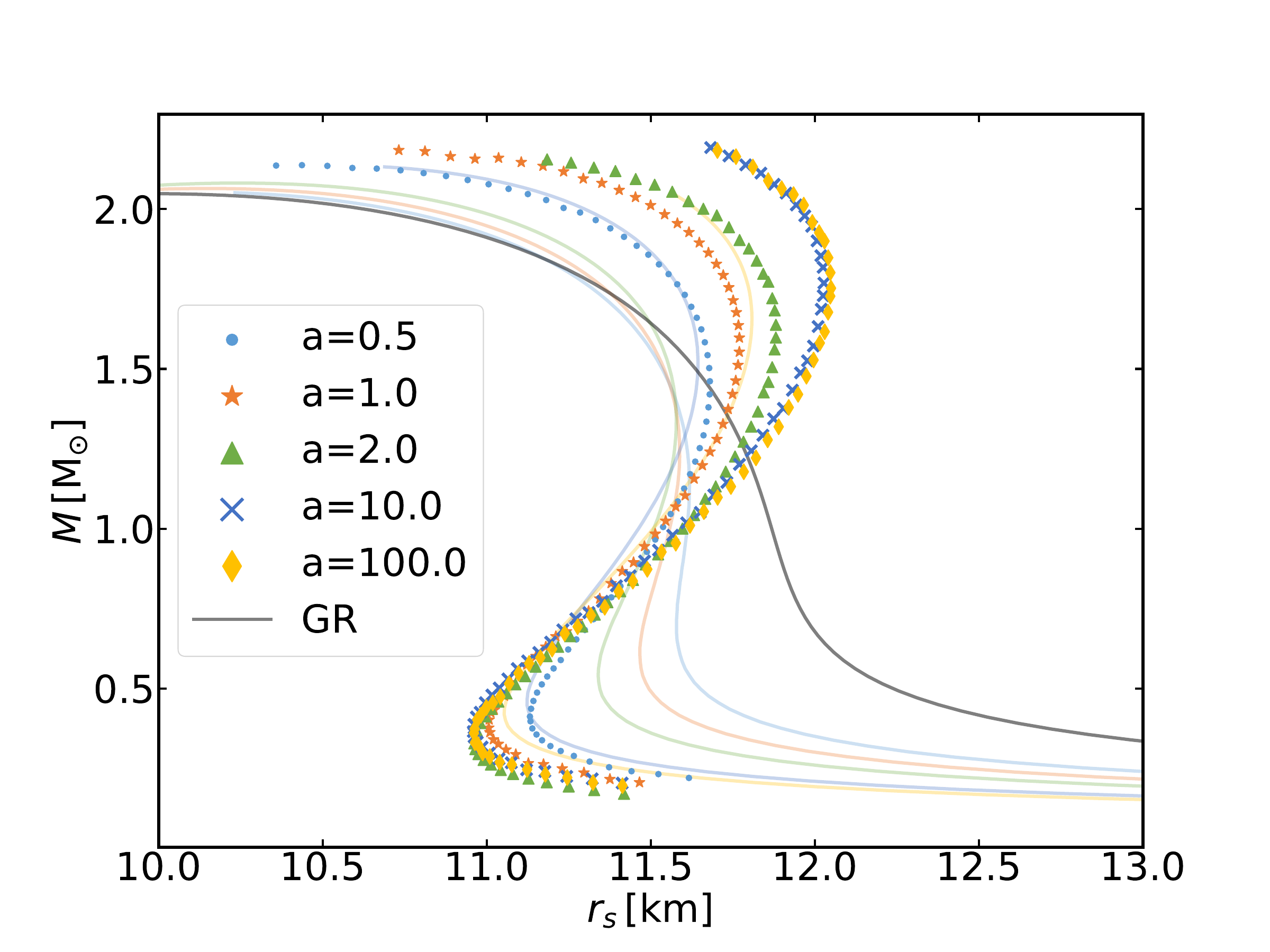}%
        \subcaption{
            Mass-$r_s$ relation 
            with $p_c$ constrained by $U_{\mathrm{eff}, \Tilde{\Phi}\Tilde{\Phi}}|_{\Tilde{\Phi} = \Tilde{\Phi}_{\mathrm{sol}}}>0$
        }%
        \label{fig:m-r_relation_for_phi_sol}%
    \end{minipage}%
    &
    \begin{minipage}{0.5\linewidth}%
        \centering%
        \includegraphics[keepaspectratio, width=0.9\linewidth]{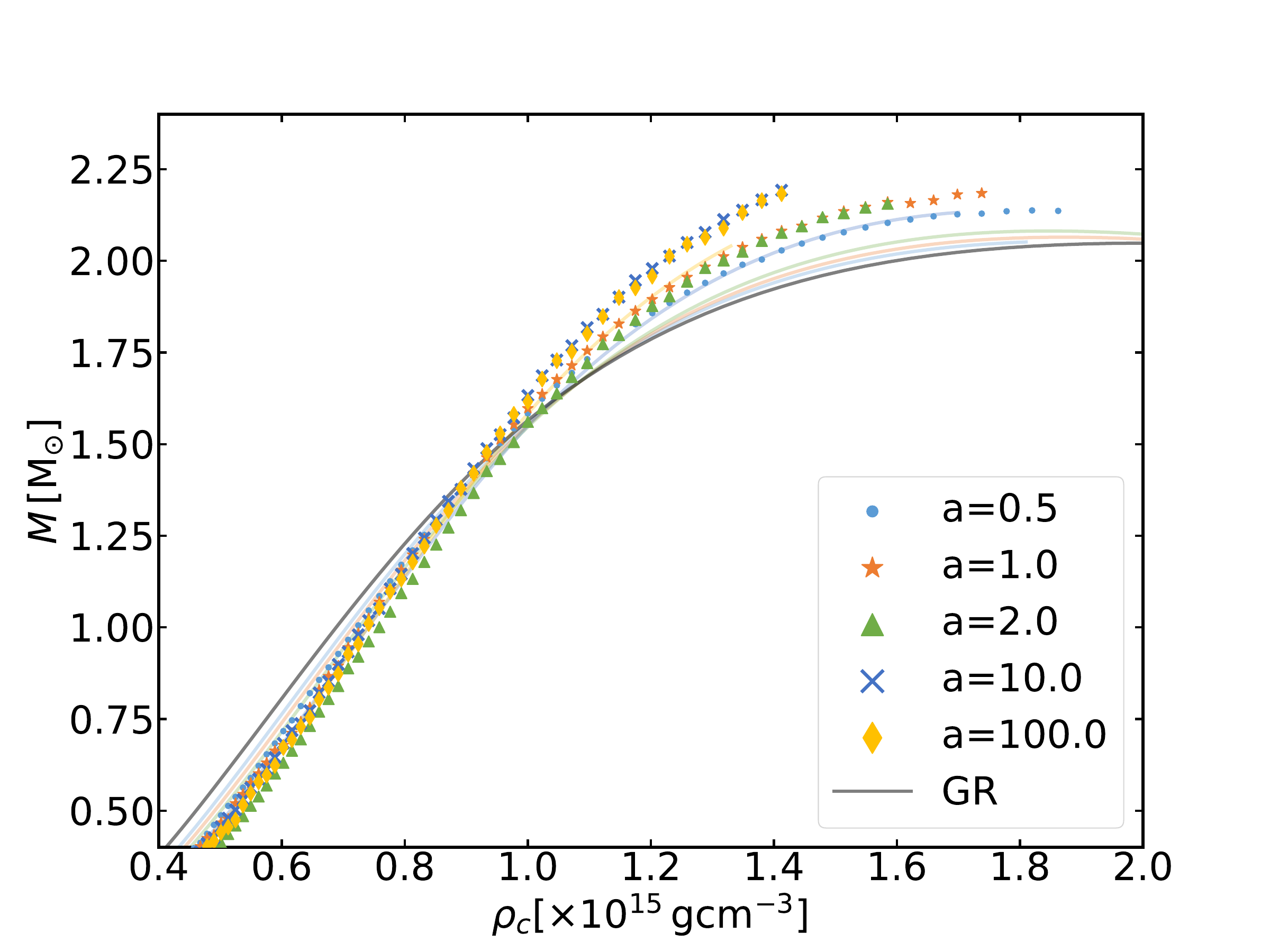}%
        \subcaption{
            Mass-$\rho_c$ relation
            with $p_c$ constrained by $U_{\mathrm{eff}, \Tilde{\Phi}\Tilde{\Phi}}|_{\Tilde{\Phi} = \Tilde{\Phi}_{\mathrm{sol}}}>0$
        }%
        \label{fig:m_rho_relation_for_phi_sol}%
    \end{minipage}%
\end{tabular}
\caption{
    The M-R relation and M-$\rho_c$ relation with the different upper limits on $p_c$ applied.
    We have chosen the same parameters as in \figref{fig:Mass-radius-density}.
    The solid curves represent the M-R relation in $R^2$ gravity as the reference.
}
\label{fig:Mass-radius-density-phi}
\end{figure}%

For the NIP gravity, the chameleon potential in the Einstein frame is derived as
\begin{align}
\label{eq:chame_pot_Einstein}
    U_{\mathrm{\mathrm{eff}}} (\Tilde{\Phi}, T)
    = 
    \frac{1}{8\kappa^{2}}e^{-2\sqrt{\frac{2}{3}}\kappa\tilde{\Phi}}
    \left[
        4\left(\frac{b}{a}\right)^{b}
        \left(\frac{e^{\sqrt{\frac{2}{3}}\kappa\tilde{\Phi}}-1}{b+1}\right)^{b+1}
        -2\kappa^{2}T
    \right]
    \, ,
\end{align}
up to a constant term. 
The plots for several $\alpha$ and $T$ are shown as \figref{fig:chame_pot_EI}. 
The second derivative of the effective potential is given by
\begin{align}
\begin{split}
    \frac{d^{2} U_{\mathrm{eff}}}{d \Tilde{\Phi}^2}
    &=
    \frac{1}{3}e^{-2\sqrt{\frac{2}{3}}\kappa\Tilde{\Phi}} 
    \\
    &\times 
    \left \{
        \left(\frac{b}{a}\right)^{b}
        \left(e^{\sqrt{\frac{2}{3}}\kappa\Tilde{\Phi}}-1\right)^{b-1} 
        \frac{ e^{\sqrt{\frac{2}{3}}\kappa\Tilde{\Phi}}
        \left[(b-1)^{2}e^{\sqrt{\frac{2}{3}}\kappa\Tilde{\Phi}}+3b-5\right] +4}
        {(b+1)^{b+1}}-2\kappa^{2}T
    \right \}
    \, .
\end{split}
\end{align}
The nontachyonic condition of the scalaron field, $m_{\Tilde{\Phi}}^2>0$, leads to the constraint on the energy-momentum tensor in the Jordan frame:
\begin{align}
    T < \frac{1}{2\kappa^{2}}\left(\frac{b}{a}\right)^{b}
    \left( e^{\sqrt{\frac{2}{3}} \kappa \tilde{\Phi}_{\min}} -1 \right)^{b-1}
    \frac{ e^{\sqrt{\frac{2}{3}} \kappa \tilde{\Phi}_{\min}}
    \left[(b-1)^{2}e^{\sqrt{\frac{2}{3}} \kappa \tilde{\Phi}_{\min}}+3b-5\right] + 4}
    {(b+1)^{b+1}} 
    \, .
    \label{eq:cond_chamemass}
\end{align}
Here, $\tilde{\Phi}_{\min}$ is determined by the stationary condition.

\begin{figure}[tbp]%
\begin{tabular}{cc}
    \begin{minipage}[t]{0.5\linewidth}%
        \centering%
        \includegraphics[keepaspectratio, width=0.9\linewidth]{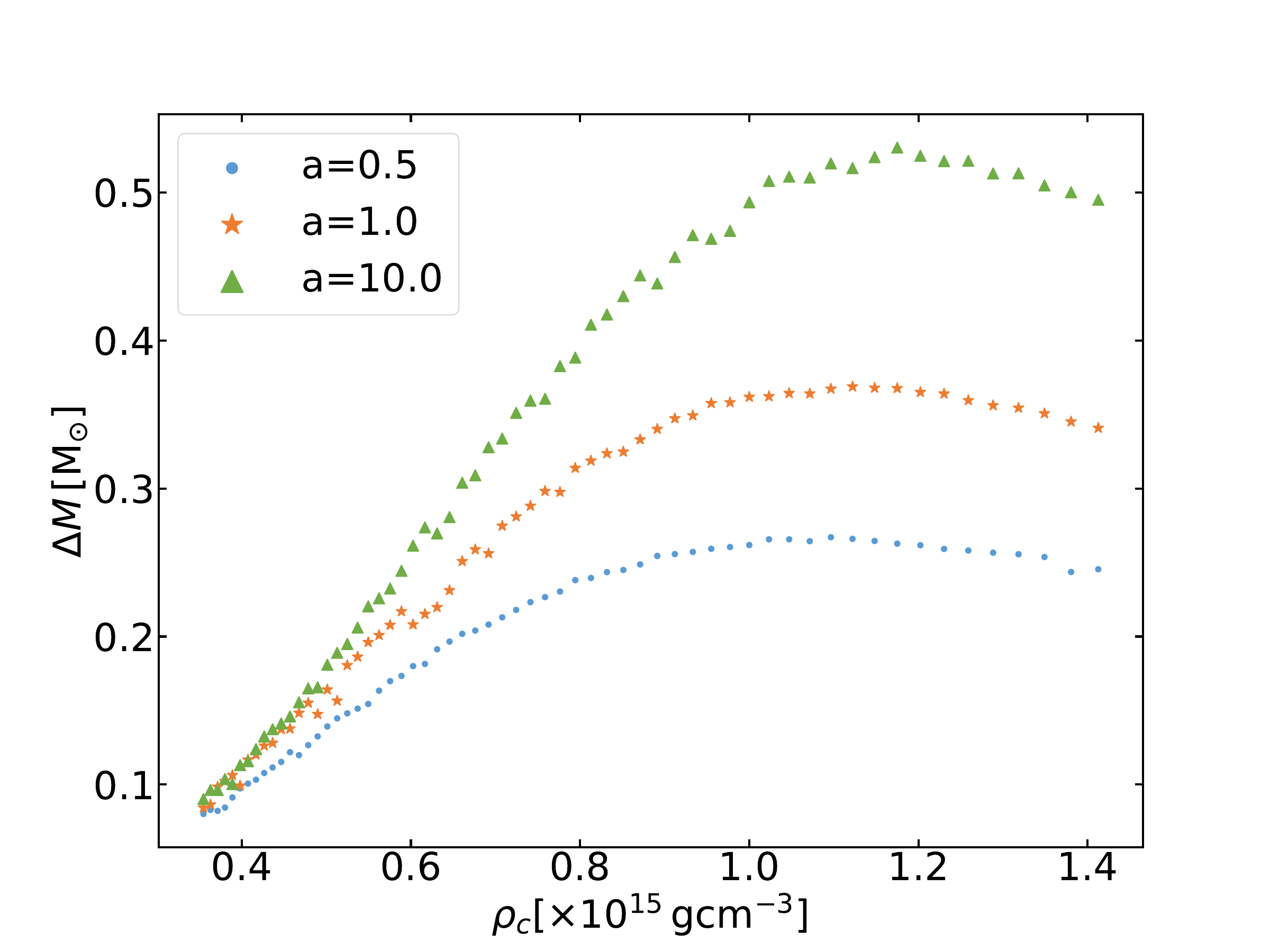}%
        \subcaption{
            $\Delta M$ with $p_c$ constrained by $m_{\Tilde{\Phi}}^2>0$
        }%
        \label{fig:Delta M for Phi_min}%
    \end{minipage}%
    &
    \begin{minipage}[t]{0.5\linewidth}%
        \centering%
        \includegraphics[keepaspectratio, width=0.9\linewidth]{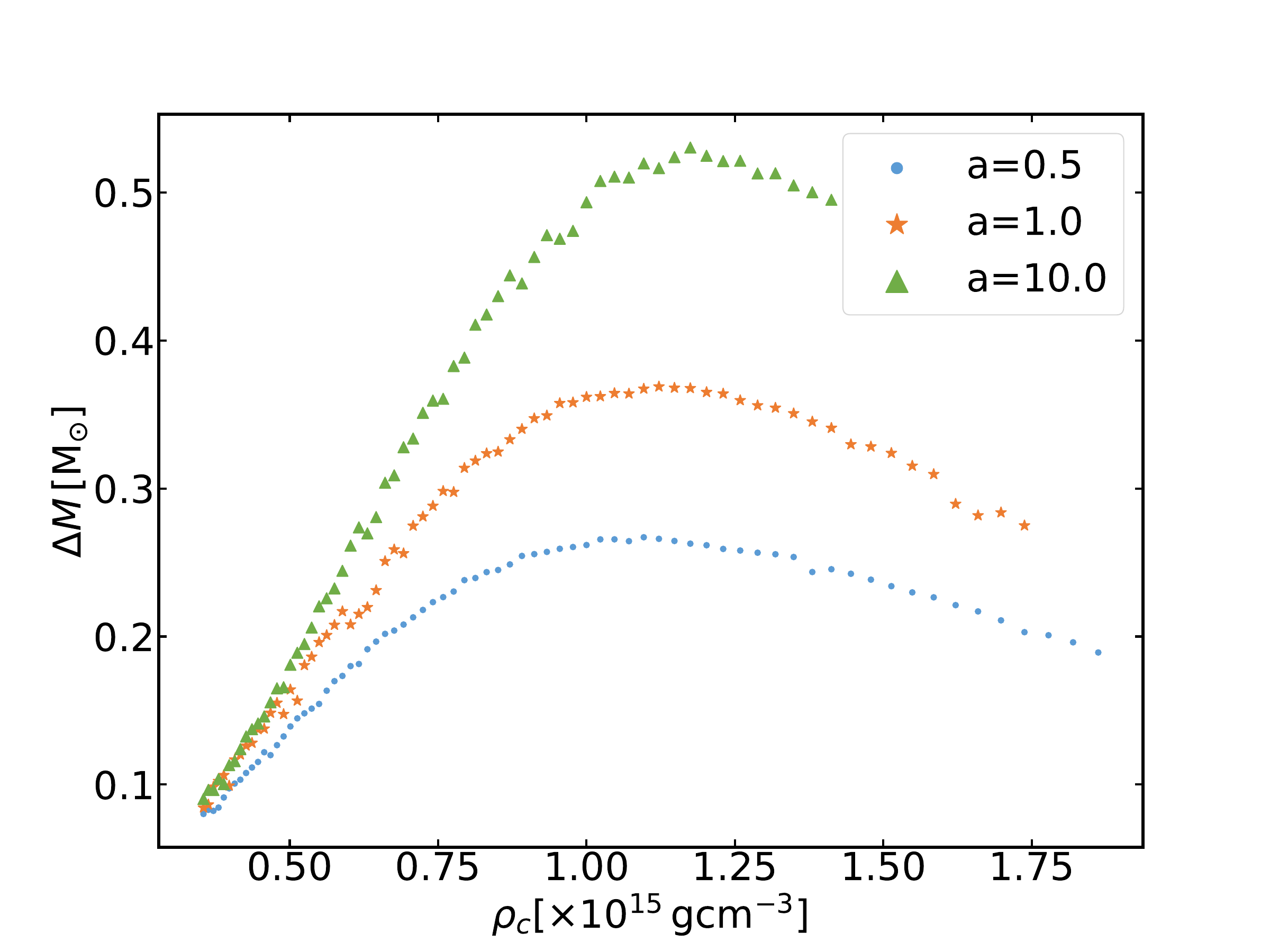}%
        \subcaption{
            $\Delta M$ with $p_c$ constrained by $U_{\mathrm{eff}, \Tilde{\Phi}\Tilde{\Phi}}|_{\Tilde{\Phi} = \Tilde{\Phi}_{\mathrm{sol}}}>0$
        }%
        \label{fig:Delta M for Phi_sol}%
    \end{minipage}%
\end{tabular}
\caption{
    The difference between $M$ and $m(r_s)$ with the different upper limits on $p_c$ applied.
    We have chosen the same parameters as in \figref{fig:Delta M}.
}
\label{fig:Delta M2}
\end{figure}%

The above constraint on $T$ gives another upper limit on the possible value of central pressure $p_c$ or energy density $\epsilon_c$ in addition to the constraint from the speed of sound $v^2$.
We can consider two different cases of the upper limit.
The first one is the nontachyonic condition $m_{\Tilde{\Phi}}^2>0$,
and the second one is the positivity for the curvature of the effective potential $U_{\mathrm{eff}, \Tilde{\Phi}\Tilde{\Phi}}>0$ at $\Tilde{\Phi}_{\mathrm{sol}}$.
Here, we can translate $\Phi_{\mathrm{sol}}$ into $\Tilde{\Phi}_{\mathrm{sol}}$ by Eq.~\eqref{eq:scalarontrans}.
The above also corresponds to the comparison between $\Phi_{\min}$ and $\Phi_{\mathrm{sol}}$ in analyzing the effective mass in \secref{sec:results}.

We plot the M-R and mass-central density relations with new upper limits on $p_c$.
We apply the same parameter choices as in \secref{sec:results}.
Comparing with \figref{fig:Mass-radius-density},
we find that M-R curves in the large-mass region are truncated due to a more severe upper limit on $p_c$ in \figref{fig:Mass-radius-density-phi}.
The only difference between them is whether the star whose central rest-mass density is very large exists or not. 
For instance, the stars whose central density $\rho_c$ is greater than $10^{15.16}\mathrm{g\,cm^{-3}}$ are not plotted in Figs.~\ref{fig:m-r_relation_for_phi_min} and~\ref{fig:m_rho_relation_for_phi_min}.
If the $\Tilde{\Phi}_{\mathrm{sol}}$ is used to constrain $p_c$,
some stars whose central density $\rho_c>10^{15.16}\mathrm{g\,cm^{-3}}$ can exit.
However, this constraint will be more strict for larger $a$.
The same results can be found in $\Delta M$ shown in \figref{fig:Delta M2}.


\bibliography{reference}

\end{document}